\documentclass[12pt, preprint]{aastex}
   \def\fdg{\hbox{$.\!\!^\circ$}}
  \begin{document}
\title{The Inclination of the Soft X-ray Transient A0620--00\\ and the Mass of its Black Hole}
\author{Andrew G. Cantrell \and Charles D. Bailyn} 
\affil{Department of Astronomy, Yale University. PO box 208101, New Haven, CT, 06520}
\email{andrew.cantrell@yale.edu, charles.bailyn@yale.edu} 
\author{Jerome A. Orosz}
\affil{Department of Astronomy, San Diego State University, 5500 Campanile Drive, San Diego, CA 92182-1221}
\email{orosz@sciences.sdsu.edu}
\author{Jeffrey E. McClintock}
\affil{Harvard-Smithsonian Center for Astrophysics, 60 Garden Street, Cambridge, MA 02138}
\email{jmcclintock@cfa.harvard.edu} 
\author{Ronald A. Remillard}
\affil{Kavli Institute for Astrophysics and Space Research, MIT, Cambridge, Massachusetts 02139}
\email{rr@space.mit.edu}
\author{Cynthia S. Froning} 
\affil{Center for Astrophysics and Space Astronomy, University of Colorado, 593 UCB, Boulder, CO 80309-0593}
\email{Cynthia.Froning@Colorado.EDU} 
\author{Joseph Neilsen} 
\affil{Harvard University Department of Astronomy, 60 Garden Street, MS-10 Cambridge, MA 02138, USA}
\email{jneilsen@cfa.harvard.edu} 
\author{Dawn M. Gelino} 
\affil{ NASA Exoplanet Science Institute, California Institute of Technology, MS 100-22, 770 South Wilson Avenue, Pasadena, CA 91125}
\email{dawn@ipac.caltech.edu} 
\author{Lijun Gou}
\affil{Harvard-Smithsonian Center for Astrophysics, 60 Garden street, Cambridge, MA 02138, USA}
\email{lgou@cfa.harvard.edu}

\begin{abstract}
We analyze photometry of the Soft X-ray Transient A0620-00 spanning nearly 30 years, including previously published and previously unpublished data.  Previous attempts to determine the inclination of A0620 using subsets of these data have yielded a wide range of measured values of $i$.  Differences in the measured value of $i$ have been due to changes in the shape of the light curve and uncertainty regarding the contamination from the disk.  We give a new technique for estimating the disk fraction and find that disk light is significant in all light curves, even in the infrared.  We also find that all changes in the shape and normalization of the light curve originate in a variable disk component.  After accounting for this disk component, we find that all the data, including light curves of significantly different shapes, point to a consistent value of $i$.  Combining results from many separate data sets, we find $i=51\fdg0\pm0.9$, implying $M=6.6\pm0.25M_\odot$.  Using our dynamical model and zero-disk stellar VIH magnitudes, we find $d=1.06\pm0.12$kpc.  Understanding the disk origin of non-ellipsoidal variability may assist with making reliable determinations of $i$ in other systems, and the fluctuations in disk light may provide a new observational tool for understanding the three-dimensional structure of the accretion disk.
  
\end{abstract}

\keywords{X-rays: binaries  ---  stars: individual (A0620-00) --- accretion disks}

\maketitle

\section{Introduction}

The prototypical soft X-ray transient A0620--00 (hereafter A0620) was discovered in its 1975 outburst, when it reached nearly 50 Crab and became the brightest X-ray nova ever observed (Elvis et al. 1975).  In quiescence, the X-ray luminosity is $~3\times10^{30}$ erg/s (Garcia et al. 2001), a tiny fraction of the stellar flux.  Optical flux in quiescence is dominated by the secondary star, a K3-7 dwarf (Froning, Robinson \& Bitner 2007 and references therein).  The mass function was first measured by McClintock \& Remillard (1986, henceforth MR86), who found $f(M)=3.18\pm 0.16 M_{\odot}$.  This number makes A0620 a strong black hole candidate and has been confirmed and refined by Marsh, Robinson \& Wood (1994, henceforth MRW94), Orosz et al. (1994), and Neilsen, Steeghs \& Vrtilek (2008, henceforth NSV08).  MRW94 and NSV08 also presented Doppler tomography of the accretion disk, revealing an asymmetrical disk with a prominent bright spot at the stream-disk impact point.  The quiescent disk has been found to vary photometrically on a range of timescales (e.g. MRW94, NSV08), and to be significant even in the infrared (Froning, Robinson \& Bitner, 2007, henceforth FRB07).  The optical variability of the X-ray quiescent disk can be categorized in three distinct states, which are distinguished by brightness, color, and aperiodic variability (Cantrell et al. 2008, henceforth CBMO08).

A0620's mass function makes it a strong black hole candidate, but a precise determination of the mass of the primary depends on knowing the inclination of A0620's orbit.  Many attempts have been made to determine the inclination by fitting ellipsoidal light curves, but the results have been inconsistent: Haswell et al. (1993) found $i>62^\circ$;  Shahbaz et al. (1994) found $i=36\fdg7$; Gelino, Harrison \& Orosz (2001, heceforth GHO01) found $i=40\fdg8$;  Froning \& Robinson (2001, henceforth FR01) found a wide range of inclinations in different epochs of data and concluded that  $38^\circ<i<75^\circ$.  These measurements of $i$ allow for masses ranging from 3.4 $M_\odot$ for $i=75^\circ$ to 13.2 $M_\odot$ for $i=37^\circ$ (MRW 1994).  In this paper, we give a comprehensive reanalysis of new and previously published light curves and find that all available data are consistent with a single value of $i$.

The disagreement between published values of $i$ has at least two origins, both of which have been discussed in the literature: secular changes in the shape of the light curve, and uncertainty in the contribution of the disk.  We now describe each of these briefly.

(1) A0620's light curve has changed shape repeatedly over the years (e.g. Leibowitz, Hemar, \& Orio 1998, henceforth LHO98), and attempts to model the different light curves have not produced consistent results.  For example, FR01 modeled three different light curves using a model with an accretion disk and hot spot, and found inclinations of 53, 70, and 74 for the different curves.  At least some of the changes in curve shape are due to optical state changes identified by CBMO08.  In what CBMO08 call the ``passive state", A0620 showed a consistent curve shape with no aperiodic variability from 1999-2003.  Other states are brighter, bluer, and show strong aperiodic variability on many timescales.

(2) The inclination derived from a given light curve depends sensitively on the disk fraction assumed in making the fit.   For a single light curve, FR01 find inclinations ranging from  from $i=44^\circ$ to $i=64^\circ$ by varying the disk fraction (defined as disk flux over total flux) from 0.2 to 0.5.  Some authors have argued that the disk is negligible in the NIR and thus assumed a zero disk fraction in determining $i$ (e.g. GHO01).  However, FRB07 find a significant NIR disk fraction and instead determine $i$ by using earlier light curves and assuming a constant disk fraction.  Subsequent work has shown that the disk fraction is highly variable: NSV08 find that the V-band disk fraction ranges from  0.48 to 0.76 over three nights of observations, and CBMO08 show that the infrared and optical show similar photometric variability.  Determining $i$ therefore requires a better understanding of variations in the disk fraction.

In this paper, we address the issues of variable curve shape and uncertain disk fractions and present a comprehensive reanalysis of light curves spanning almost 30 years.  We find that once these two problems are addressed, the full data set indicates a self-consistent value of $i$.  In Section 2, we describe the available data and select a subset which will give the most reliable determination of $i$.  In Section 3, we present a method for determining the disk fraction in specific curves, and we derive the disk fraction for each of the good light curves identified in Section 2.  In Section 4, we fit ellipsoidal models to these light curves, fixing the disk fraction using the results of Section 3.  Fitting each curve individually, we find that many curves give consistent determinations of $i$; combining these determinations of $i$, we find $i=50\fdg98\pm0.87 $, implying $M=6.6\pm 0.25 M_\odot$.  In Section 5, we use our firm dynamical model to make an improved estimate of the distance to A0620.  In Section 6, we discuss our results, including the implications of this work for future determinations of $i$ using ellipsoidal variations in other systems. 

\section{Collection of Passive-State Light Curves}

CBMO08 identified three distinct optical states in X-ray quiescence and argued that only data in what they call the passive state may be reliably used to determine the inclination.  Whereas passive-state data in CBMO08 retain a consistent curve shape across 4 years of observations, the active state data show a variable and poorly-defined curve shape.  In addition, active-state data have a large and variable disk fraction: In active state optical observations on 2006 December 14, NSV08 observed the V-band disk fraction to change from 59\% to 75\% in just 10 minutes (J. Neilsen, private communication).  Such large fluctuations in the disk fraction pose a serious challenge to a determination of $i$ from ellipsoidal light curves; the problems with using active state data to determine $i$ will be discussed in detail in Section 6.  We therefore carry out a reanalysis of existing data, using exclusively passive state data to determine $i$. 

In this section, we present a comprehensive collection of passive state light curves with well-defined curve shapes.  In Section 2.1, we describe the full photometric data set considered in this paper and describe the calibration procedures used to combine curves from different epochs.  In Section 2.2, we restrict our attention to passive-state data, which we identify using a more general definition than that given in CBMO08.  Within the passive state data, we identify a limited set of curves which show a consistent light curve over at least a full orbit, and are therefore usable for determinations of $i$.  

\subsection{Collection and Reduction of Data}

The data considered in this paper were obtained by many observers between 1981 and 2008, and include most previously published light curves and some previously unpublished data; these data are summarized in Table 1.  Our analysis includes the full data sets presented in MR86, LHO98, FR01, GHO01, and CBMO08.  Details of the observations and reduction of these data sets may be found in the respective papers.  Though some of these data sets have previously been published only as folded light curves binned in phase, we were able to obtain unbinned versions of all data sets except GHO01.  Unbinned data are necessary to reliably identify passive states and changes in curve shape, so we will postpone an analysis of the GHO01 data until Section 4.  

In order to compare data from different papers, we use common comparison stars to perform a magnitude calibration between different data sets.  MR86 includes both I-band data and data obtained using a Corning 9780 filter, which they refer to as ``B+V" and which we call the W-band.  The W-band filter has a FWHM covering 3740-5750\AA\, comparable to the total range covered by the Johnson B and V filters.  MR86 give results in the W- and I-bands as a ratio of A0620's brightness to that of a single comparison star, which they call star T.  In the I-band, we convert these ratios to differential magnitudes, $(I_A-I_T)$, where subscripts $A$ and $T$ denote the magnitudes of A0620 and star T, respectively.  We determine $I_T$ using SMARTS I-band photometry and add this to the differential magnitudes to get $I_A$.  We were unable to perform any color correction, as we have no I-band comparison stars other than star T: McClintock \& Remillard (1983) and MR86 include comparison stars in the W-band only.  The MR86 I-band magnitude calibration could therefore be significantly off if the I-band filter used in MR86 differs from the SMARTS filter.

We normalize the W-band data so that its magnitude calibration is consistent with the SMARTS V-band data; doing so requires a magnitude calibration and color correction.  We do not attempt to adjust the shape of the curve to agree with V-band curves: the W-band curve shapes still reflect the W-band filter, but we adjust their normalization so that the overall brightness of the data may be compared directly to V-band data.  We begin by converting the ratios given in MR86 to differential W-band magnitudes, $(W_A-W_T)$.  We then compute a differential color correction (described in the next paragraph) to convert these values to differential V-band magnitudes, $(V_A-V_T)$.   Adding the SMARTS value of $V_T$ then gives $V_A$, so that the MR86 W-band data have the same normalization as the SMARTS V-band data.  We believe that this magnitude calibration is more reliable than the I-band calibration, since we had multiple comparison stars and could therefore determine a color correction.

To determine the differential color correction relating $(W_A-W_T)$ to $(V_A-V_T)$, we use three field stars with differential photometry in MR86, giving us $(W_*-W_T)$ for each star.  Using SMARTS V-band magnitudes, we find ${(W_*-W_T)-(V_*-V_T)}$ for each star and plot these values as a function of V-I color.  We then fit a line and determine the value of ${(W_*-W_T)-(V_*-V_T)}$ corresponding to the V-I color of A0620.  We find $${(W_A-W_T)-(V_A-V_T)=0.0\pm0.1}.$$  The color correction is negligible for two reasons: (1) star T has almost the same color as A0620 (V-I=1.48 for star T and 1.54 for A0620), so the differential magnitude depends only weakly on the filter, and (2) ${(W_*-W_T)-(V_*-V_T)}$ is a very weak function of V-I, with the three stars used in McClintock \& Remillard (1983) giving statistically consistent values of $(W_*-W_T)-(V_*-V_T)$.  In particular, though the color of A0620 is known to be slightly variable (FR01), these small changes in color will not affect the magnitude calibration.  Though our color correction is based on only three stars, we believe it is trustworthy: point (1) would be true even if more stars were available, and point (2) is unlikely to be qualitatively wrong, given the overlap between the filters and the fact that for red objects like A0620, W-band flux will be dominated by emission in the V-portion of the ``B+V" filter.  We therefore believe the essential conclusion is robust and that the differential color correction is negligible. 

Similarly, we use common comparison stars to give a common magnitude calibration to the FR01 H-band data and the SMARTS H-band data.  Specifically, we compute for each observation in FRB01 the differential magnitude between A0620 and the mean of the comparison stars called "star 1" and "star 2" in Table 2 of FR01.  We then add the mean calibrated magnitude of these two stars in the SMARTS images, to obtain a common magnitude calibration.  However, this magnitude calibration is unreliable, as the comparison stars listed in FR01 do not give a self-consistent calibration relative to the SMARTS data.  For example, their star 1 and star 2 differ by 1.43 mag, while they differ by 1.48 mag in the SMARTS images.  With only three comparison stars to work with, it is impossible to identify the origin of this discrepancy, or estimate its impact on the calibration of A0620.

In addition to the previously published data discussed above, we include two data sets obtained by JEM and RAR, that have not been presented elsewhere.  The first of these was obtained using the McGraw-Hill 1.3m during 1987--1992 (W- and I-band).  These data are an extension of the photometry published in MR86 and were gathered and reduced following the procedures described therein.  We apply the same magnitude calibration and color correction used on data from MR86, so that all W- and I-band data may be compared directly to the V- and I-band data in other papers; as in the MR86 I-band data, we were unable to test for any color term in the I-band magnitude calibration, making the magnitude calibration particularly uncertain.  

The second of the previously unpublished  data sets was obtained by JEM and RAR using the Fred Lawrence Whipple Observatory (FLWO) 1.2m during 1993-1996 (W-, V- and I-band).  The FLWO data were reduced using standard IRAF tasks. To give a consistent magnitude calibration across all I-band data, aperture photometry was performed using the same comparison stars and magnitude calibration as CBMO08.   

All data are folded on the ephemeris given in Johannsen, Psaltis, \& McClintock (2009), which is sufficiently accurate to determine $T_0$ to within 0.005 in phase throughout the 27 years spanned by our data.  We believe that our magnitude calibrations are robust, with the possible exceptions discussed above, namely JEM and RAR's McGraw-Hill I-band data and the FR01 H-band data.  We therefore have a consistent phasing and magnitude calibration for a wide range of data sets, allowing these data to be combined and compared.  

\begin{deluxetable}{llllc}
\tablecaption{Photometric Data Sets\tablenotemark{a}}
\tabletypesize{\footnotesize}
\tablehead{
\colhead{Data Set  \tablenotemark{b}} & \colhead{Date Range} & \colhead{Filters\tablenotemark{c}} & \colhead{$n$\tablenotemark{d}} & \colhead{Number of Nights}}

\startdata
MR86 & 1981 Oct 29-1985 Jan 17 & WI & 360 & 23\\
MR1 & 1987 Jan 1-1991 Feb 7 & WI & 891 & 11\\
LHO98 & 1991 Jan 08-1995 Nov 19 & R & 953 & 28 \\
MR2 & 1993 Jan 19-1996 Jan 15 & WVI  & 349 & 10\\
FR01 & 1995 Dec 15-1996 Dec 12 & JHK & 1479 & 11\\
GHO01\tablenotemark{e} & 1999 Feb 25-2000 Dec 11 & JHK& -- & 6\\
CBMO08 & 1999 Sept 11-2007 April 1& VIH & 4816 & 934\\
\enddata

\tablenotetext{a}{\ List of all data sets used in this paper; only a fraction of this data is selected for use in determining $i$.  See Section 2.2 for the selection criteria applied.}

\tablenotetext{b}{MR1=Data obtained by McClintock \& Remillard at the McGraw Hill 1.3m, extending the data published in MR86; MR2=Data Obtained by McClintock \& Remillard at FLWO, 1993-1996.  All other data sets are referenced by the paper in which they first appeared.  See text for details.} 

\tablenotetext{c}{See Section 2.1 for a description of the W-band filter; all others are standard.}

\tablenotetext{d}{Total number of exposures in all bands.}

\tablenotetext{e} {We were unable to obtain an unbinned version of the GHO01 data set, so we do not know the total number of exposures in the set.  The analysis of Section 2.2 cannot be performed on binned data, so we postpone a discussion of the GHO01 data set until Section 4.}

\end{deluxetable}

\subsection{Selection of Curves for Use in Determining $i$}

In this section, we identify light curves which may reliably be used to determine $i$.  We begin by expanding the definition of passive given by CBMO08, so that we may identify passive data in other data sets.  We then identify all the passive nights in the data described in Section 2.1.  Even among passive-state light curves, we find that there is significant variation in the shape of A0620's light curve.  We show that combining curves of different shapes may produce spurious results, and argue that only curves with a verifiably consistent shape and normalization should be used.   We therefore apply a second cut, eliminating light curves which lack compete phase coverage of a single shape and normalization.  After these two cuts, we are left with 12 passive-state, single-shape, single-normalization light curves, which are the most suitable for determining $i$. 

Our first cut, identifying passive-state data, is made based on brightness, color, and aperiodic variability.  For the SMARTS data, obtained nightly over a period of many years, we use the same selection criterion as in CBMO08: we identify passive periods based on a 10-day running average of V and V-I: any 10 day period with a phase-adjusted average $V<16.57$ or $(V-I)<1.52$ is identified as non-passive, while all other periods are considered passive; see CBMO08 for details.   For data sets taken with a more traditional cadence than the SMARTS data, a 10-day running average is not practical for the identification of passive data.  For these data sets, we use the fact that non-passive data show substantial aperiodic variability, unlike their passive-state counterparts (CBMO08).  For all data other than the SMARTS data, we therefore identify as passive any night which shows negligible aperiodic variability and which, at some phase, lies near the lower envelope of data in a given band.  Specifically, a night is considered passive if it meets two criteria: (1) Aperiodic variability, defined as the scatter of data in phase bins of width 0.03, has a standard deviation less than the photometric errors or 0.035 mag, whichever is larger; and (2) At some phase, the data are within 0.1 mag of the lower envelope of all data in the same band and at the given phase.      

Table 2 lists all the nights on which we were able to determine whether A0620 was passive, based on the two criteria above.  With the exceptions of GHO01 and LHO98, all data sets listed in Table 1 appear in their entirety in Table 2, split into passive and non-passive nights.  We were able to obtain the GHO01 data only in binned form, so we cannot reliably identify passive and active nights within their data set.  The R-band data of LHO98 consistently show aperiodic variability greater than 0.035 mag, making the entire data set non-passive by our definition.   However, there is evidence for generally enhanced variability in the R-band (Haswell et al. 1993).  We suspect that the enhanced aperiodic variability in the R-band is due to variation in the H-$\alpha$ line, rather than variability in the disk continuum flux.  Since variations in the continuum drive the state changes observed in other bands, R-band variability does not necessarily correlate with these state transitions.  It is therefore likely that A0620 was passive on some nights of LHO98 observations, but that the R-band cannot be used to identify the states apparent in other bands.  We exclude R-band data from subsequent analysis because of the high level of aperiodic variability and the uncertainty in the identification of passive nights.

\begin{deluxetable}{lcl}
\tablecaption{Identification of Passive and Active Data Sets \tablenotemark{a}}
\tabletypesize{\footnotesize}
\tablehead{
\colhead{Date Range} & \colhead{Passive?} & \colhead{data set} } 

\startdata
1981 Oct 29-Nov 03 & N & MR86\\
1982 Dec 15-20 & N & MR86\\
1983 Dec 6-16 & N & MR86\\
1985 Jan 12-17 & N & MR86\\
1987 Jan 1-4 & N & MR1\\
1988 Jan 12-15 & Y &  MR1\\
1989 Jan 12-16 & Y & MR1\\
1991 Feb 07 & N & MR1\\
1992 Jan 01 & N & MR1\\
1993 Jan 20-25  & N & MR2\\
1994 Jan 5-8 & Y & MR2\\
1995 Dec 16-18 & N & FR01\\
1996 Jan 14-16  & N & MR2\\
1996 Jan 27-29 & Y & FR01\\
1996 Dec 7-12 & Y & FR01\\ 
1999 Sept 12-22 & N & CBMO08\\
1999 Sept 23-Oct 29 & Y & CBMO08\\
1999 Oct 30-Nov 11  & N & CBMO08\\
2000 Jan 26-2000 Feb 05  & N & CBMO08\\
2000 Feb 13-April 09 & Y & CBMO08\\
2000 Oct 1-2001 Feb 20 & Y & CBMO08\\
2001 Feb 21-Mar 25  & N & CBMO08\\
2001 Mar 26-Apr 10 & Y & CBMO08\\
2001 Oct 13-2001 Nov 6  & N & CBMO08\\
2001 Nov 10- 2002 Jan 25 & Y & CBMO08\\
2002 Feb 08-2002 Mar 29 & Y & CBMO08\\
2002 Mar 31-May 09  & N & CBMO08\\
2003 Feb 04-Apr 04 & Y & CBMO08\\
2003 Oct 03-Nov 07 & Y & CBMO08\\
2003 Nov 08-Dec 10  & N & CBMO08\\
2003 Dec 12-Dec 15 & Y & CBMO08\\
2003 Dec 16-2007 Apr 1 & N & CBMO08\\

\enddata

\tablenotetext{a}{Nights on which A0620 could be reliably identified as passive or non-passive.  States could not be identified from the GHO01 data, as we obtained the data in binned form only.  States also could not be identified from the LHO98 data because of the persistently high level of R-band variability (see text).}

\end{deluxetable}

The full collection of passive-state data includes many partial nights of data with incomplete phase coverage.  It is tempting to combine these data to get complete light curves, but we now argue that this can produce misleading results.  In Figure 1, we show light curves obtained by JEM and RAR, on the nights of 1989 January 12, 14, and 15.  Though the January 14-15 data are consistent, the curve shows a clear change of both shape and normalization between January 12 and January 14.  If all three nights of data were binned together, we would get a spurious light curve with a shape dependent on the orbital phases covered each night.  In addition, if we had January 14 data only between phases 0.8 and 0.35, we might combine the three nights and obtain a smooth curve covering all phases, never realizing that the combined curve is drawn from different light curves indicating different physical conditions in the binary.  Model light curves assume that all variation is due to our changing viewing angle of the system; spurious results will thus be given by fitting curves which, like the January 12-15 curves, include variation intrinsic to the system.  

\begin{figure} \figurenum{1}
\plotone{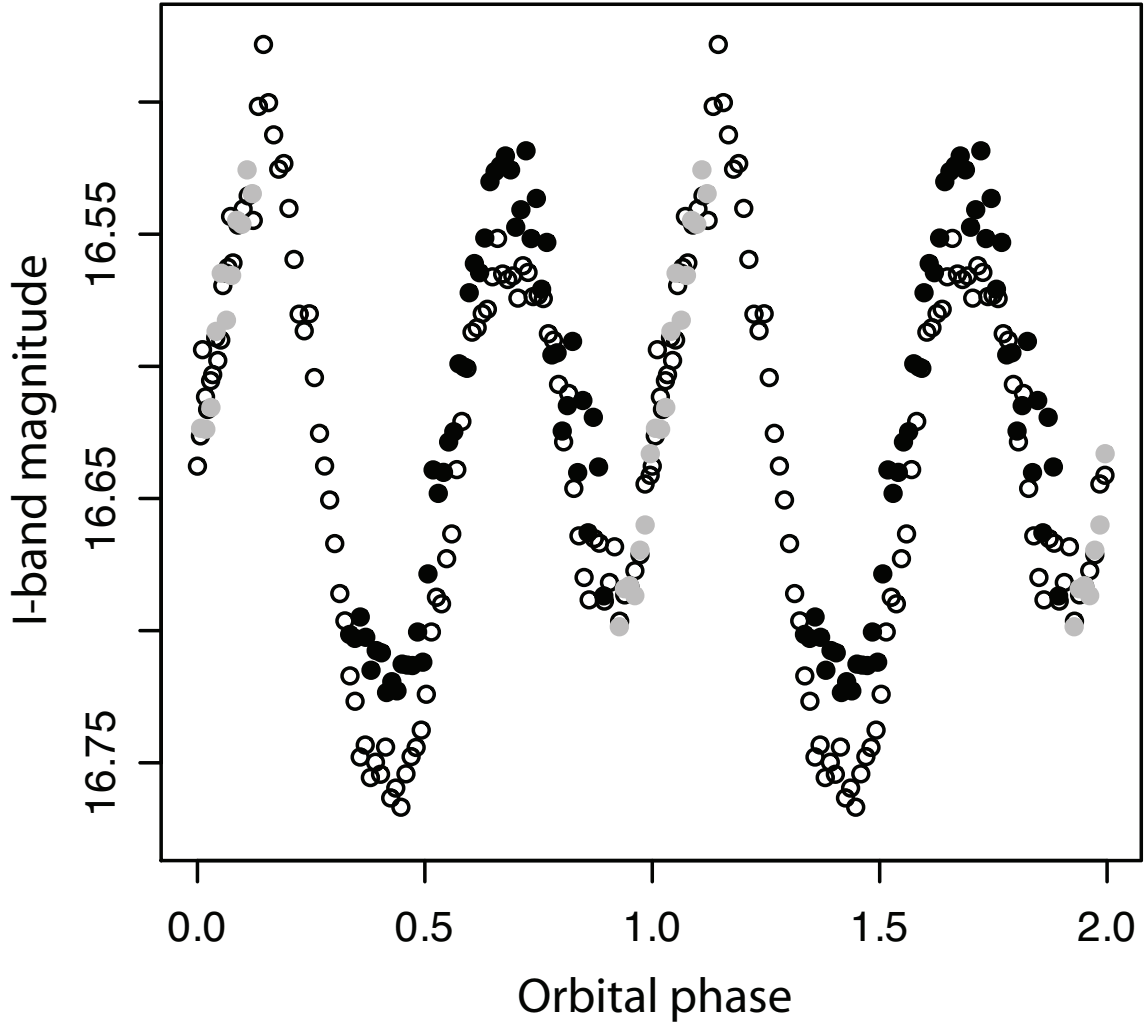}
\caption{I-band photometry of A0620.  Data were taken in 1989 on the nights of January 12 (black circles), January 14 (open circles), and January 15 (grey circles).  The data show that the light curve may change on short timescales even in the passive state, which provides a caution about combining data obtained on different nights.} 
\end{figure}

Considering the risk posed by combining data taken on different nights, we allow ourselves to combine data into a single curve only when there is clear evidence that all the data being combined comes from a single-shape, single-normalization curve.  We therefore require that the data being combined meets one of two criteria: either (1) there is a single night of observations which covers at least one full orbit, is self-consistent, and is consistent with any data combined from consecutive nights; or (2) data taken on consecutive nights cover every orbital phase at least twice and all overlaps are consistent.  For example, criterion 1 allows us to combine the January 14 and 15 data shown in Figure 1;  criterion 2 allows us to combine all passive SMARTS data, which shows a verifiably consistent shape across several years of observations even though no one night covers a full orbit.  Combining data only when permitted by one of the above criteria, we identify twelve light curves from seven epochs which are usable for determining $i$.  These light curves are listed in Table 3 and shown in Figure 2.    

The light curves listed in Table 3 include data in four wavebands from FR01, CBMO08, and previously unpublished data obtained by JEM and RAR.  Figure 2 shows substantial variability among the passive light curves, even within a single band: I-band curves range from nearly symmetric (I4), to having one very deep minimum (I6), to having uneven minima and maxima (I8).  These curves will form the basis for our determination of $i$, and the range of shapes fit by our models will serve as a consistency check on our results.

\begin{deluxetable}{llllcll}
\tablecaption{Passive, Single-Shape, Single-normalization light curves \tablenotemark{a}}
\tabletypesize{\footnotesize}
\tablehead{
\colhead{Curve ID  \tablenotemark{b}} & \colhead{Data Set \tablenotemark{c}} & \colhead{Date Range} & \colhead{$n$} & \colhead{$f_{0.554}$\tablenotemark{d}} & \colhead{$i$ \tablenotemark{e}}}

\startdata
W1  & MR1 &1988 Jan 12 & 22  & $0.34\pm0.03$ & $48.79\pm2.73$ \\
I1\tablenotemark{f}  & MR1 &1988 Jan 12 & 64  & $0.24\pm0.03$ & $53.36\pm1.66$ \\
W2  & MR1 &1988 Jan 13 & 20  & $0.34\pm0.03$ &$44.42\pm4.60$ \\
I2\tablenotemark{f}  & MR1 &1988 Jan 13 & 61  & $0.20\pm0.03$ & $48.14\pm1.20$ \\
W3 & MR1 & 1988 Jan 14 & 24 & $0.40\pm0.03$ & $52.56\pm4.31$ \\
I3\tablenotemark{f} & MR1 & 1988 Jan 14 & 71 & $0.26\pm0.03$ & $45.92\pm1.32$ \\
I4 & MR2 & 1994 Jan 4- Jan 7 & 122 & $0.19\pm0.03$ & $56.19\pm3.27$ \\
V6 & CBMO08 & 1999 Sept 23-2003 Dec 15 & 769 & $0.35\pm0.03$ & $51.75\pm1.05$\\
I6 & CBMO08 & 1999 Sept 23-2003 Dec 15 & 742 & $0.25\pm0.03$ & $50.13\pm1.35$\\
H6 & CBMO08 & 2000 Sept 30-2003 Dec 15 & 858 & $0.13\pm0.02$ & $51.58\pm3.00$\\
H7\tablenotemark{f} & FR01 & 1996 Jan 27-29, 1996 Dec 7-12 & 907 & $0.16\pm0.02$ & $43.20\pm0.81$ \\
I8 & MR1 & 1989 Jan 14-Jan 15 & 115 & $0.15\pm0.03$ & $50.92\pm1.18$ \\
\enddata

\tablenotetext{a}{A comprehensive list of passive state light curves with a consistent shape and normalization.  See Section 2.1 for a description of the full data set from which these curves were taken and Section 2.2 for details of our selection criteria.}

\tablenotetext{b}{\ Letters indicate the wavebands in which the observations were made, while numbers identify overlapping observations.  For example, W1 and I1 consist of W- and I-band observations from the same night.} 

\tablenotetext{c}{As in Table 1.}

\tablenotetext{d}{Phase 0.554 disk fractions, determined assuming that the disk is the source of all photometric variability at this phase.  Quoted errors include only the uncertainty inherited from the simultaneous spectroscopy and photometry used to determine the zero-disk magnitude in each band.  These errors exclude uncertainty in the magnitude calibration, which are discussed in Sections 2.1 and 4.  W- and V-band factions are determined in Section 3.1, H-band fractions are determined in Section 3.2 and refined in Section 3.3, and I-band fractions are determined in Section 3.4.}

\tablenotetext{e}{Inclination determined by using ELC to model the light curve.  Each light curve is modeled separately with a model including ellipsoidal variation and a spotted disk.  In each case, $f_{0.554}$ is constrained to agree with the value given in this table, to within the uncertainty given in this table.  The fits and their residuals are shown in Figure 2.  See Section 4 for details of the fits.}

\tablenotetext{f}{\ Light curves I1, I2, I3, and H7 come from data sets with particularly uncertain magnitude calibrations.  The values of $f_{0.554}$ and $i$ are therefore particularly unreliable in these curves and should be treated with caution.  See Sections 2.1 and 4.2 for details}

\end{deluxetable}

\begin{figure} \figurenum{2a}
\plotone{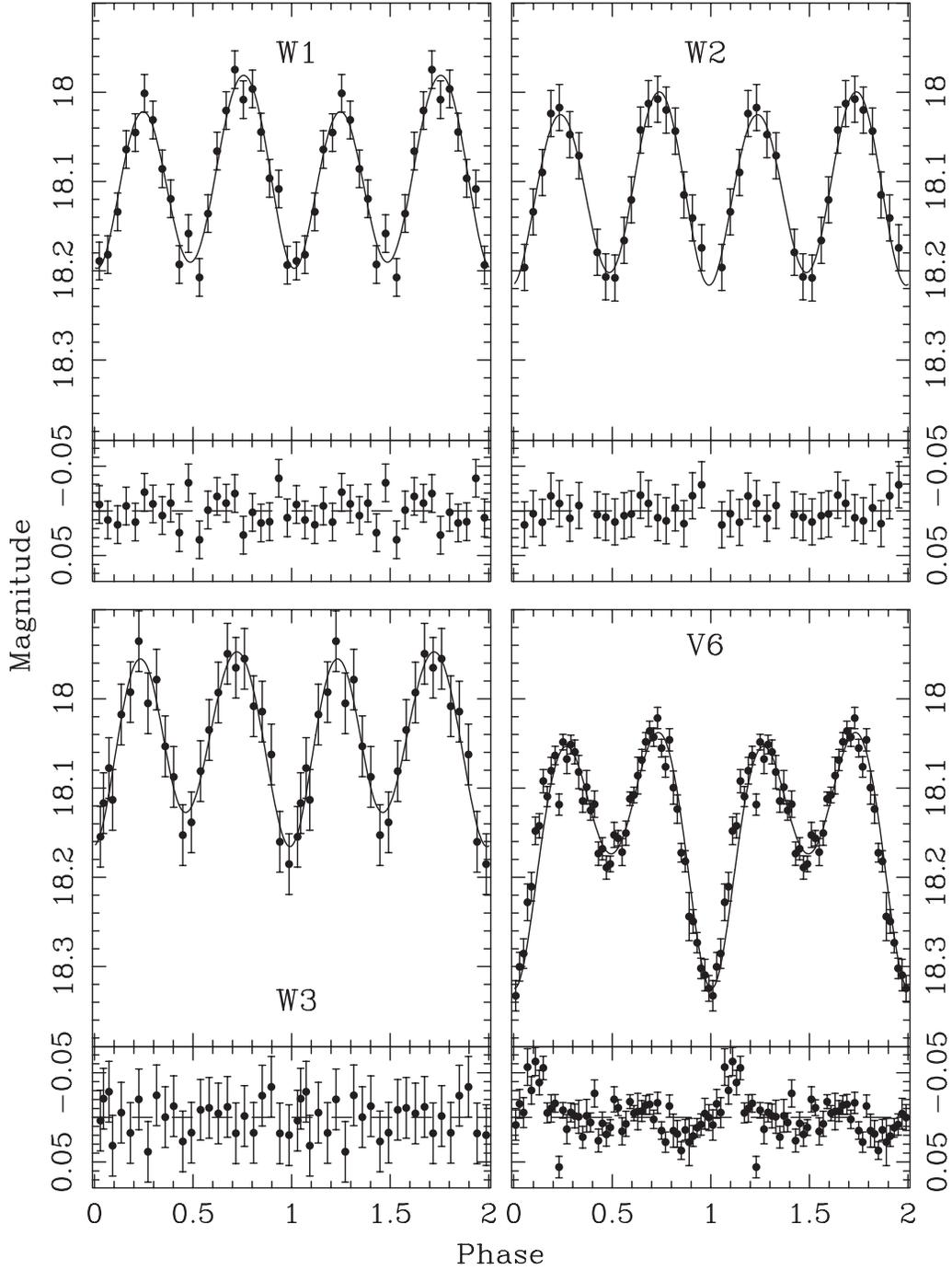}
\caption{Passive, single-shape, single-normalization light curves with full phase coverage.  Details of each curve are listed in Table 3.  To facilitate comparison, the axes are identical for all curves in a given waveband (and for V- and W-band light curves).  Unbinned data are plotted for curves with fewer than 500 data points; V6, I6, H6, and H7 are binned.  Lines are best fit models to unbinned data, with the disk fractions constrained by the values in Table 3; see Section 4 for a description of the fits and Table 3 for the implied values of $i$.  Residuals to the fits are shown beneath the respective plots.  Figure 2a: curves W1, W2, W3, and V6; Figure 2b: curves I1, I2, I3, and I6; Figure 2c: curves I4, I8, H7 and H6.  } 
\end{figure}

\begin{figure} \figurenum{2b}
\plotone{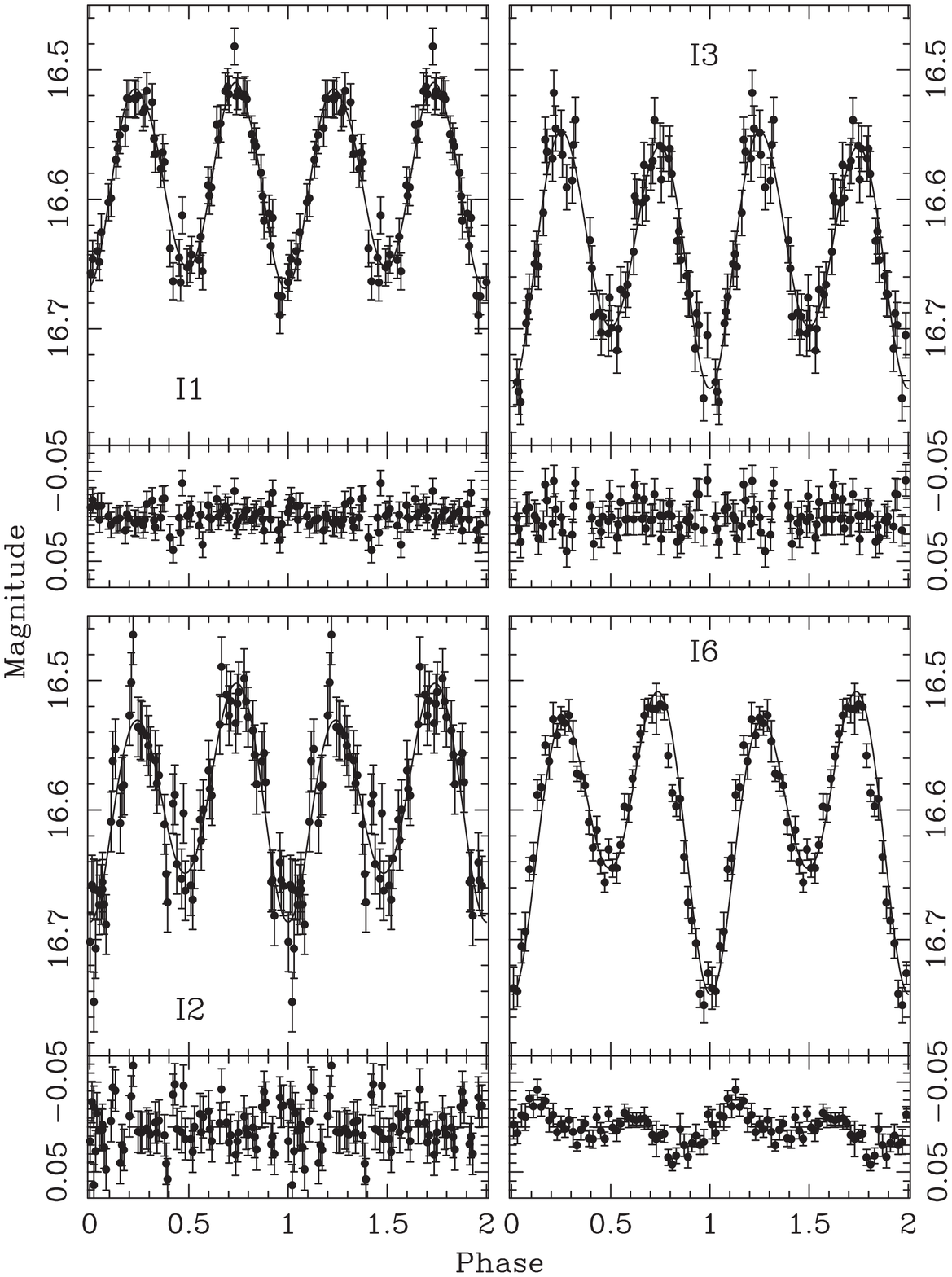}
\caption{}
\end{figure}

\begin{figure} \figurenum{2c}
\plotone{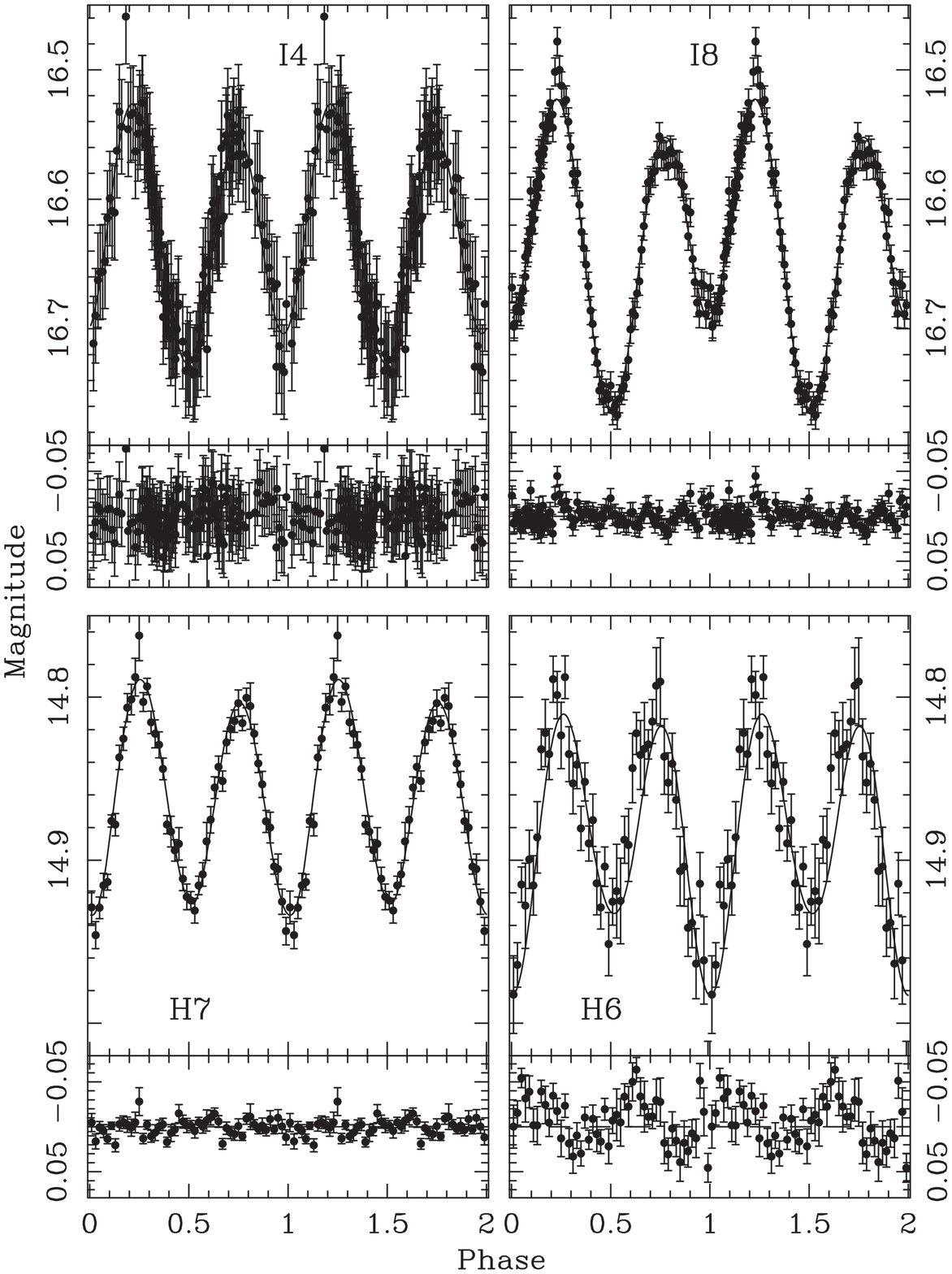}
\caption{} 
\end{figure}

\section{The Variable Disk Fraction of A0620}

The inclination derived by fitting ellipsoidal light curves depends sensitively on the disk fraction assumed in the fits (FR01),  and authors have fit light curves using various assumptions on the amount and variability of the disk light.  FRB07 found an H-band disk fraction of 18\%, demonstrating that one cannot assume the disk to be negligible, even in the infrared.  NSV08 used the veiling of stellar lines to make a time-resolved determination of the V-band disk fraction, and found that the disk fraction can change substantially on timescales ranging from minutes to days.  In addition, the broad range of brightness observed in CBMO08 suggests that this variability extends to the infrared, and occurs on timescales ranging from days to years.  Since the disk is neither negligible nor constant, a curve-by-curve determination of the disk fraction is necessary.  In this section, we use spectroscopic and photometric observations to determine the disk fraction in each light curve listed in Table 3.  

Throughout this section, we assume that the stellar flux at a given phase is constant, so that any change in brightness at a given phase is interpreted as a change in disk flux.  It is not immediately obvious that this is a valid assumption: GHO01 and others have argued that star spots dominate changes in the shape of A0620's light curve, implying that the stellar flux at a given phase is variable.  However, we will show in two ways that the disk dominates nonellipsoidal variation, with stellar flux constant at a given phase: (1) in Section 3.3, we will show that nonellipsoidal variations have a color consistent with a disk origin and inconsistent with a stellar origin, and (2) in Section 4, we show that the disk fractions derived here give a consistent value of $i$ for a wide range of curves, which provides a consistency check on the disk fractions themselves.  

In Section 3.1, we make determinations of disk fractions in W and V, extrapolating from a spectroscopic determination of the disk fraction with simultaneous photometry.  In Section 3.2, we similarly determine the H-band disk fractions.  In Section 3.3, we perform a consistency check, demonstrating that the color of nonellipsoidal variations is consistent with the disk color given by our V- and H-band disk fractions.  Finally, in Section 3.4, we determine I-band disk fractions, using a different method based on the results of Section 3.3.   We thus determine disk fractions for all curves in Table 3.

\subsection{V- and W-band Disk Fractions}

In this section, we make a determination of the disk fractions in the W- and V-band light curves listed in Table 1.  Assuming constant stellar flux at a given phase (see above), we use a spectroscopic determination of the disk fraction, together with simultaneous V-band photometry, to determine how bright A0620 would be at the given phase if all disk light disappeared.  In other V- and W-band light curves, the disk fraction at the same phase can then be determined by comparing the observed brightness to the computed zero-disk brightness.  The disk fractions {\it at this phase} are used to constrain our fits in Section 4.   

On the night of 2006 December 15, NSV08 observed a V-band disk fraction of $0.516\pm0.009$ in an 11.6-minute exposure ending at HJD 2454084.7187 (J. Neilsen, private communication).  In a 6-minute exposure beginning 27 seconds after the end of the NSV08 exposure, V-band SMARTS photometry found V=$17.83\pm0.03$.  The two exposures do not quite overlap, but the disk was stable at the time: the two subsequent spectra obtained by NSV08 give disk fractions of 0.503 and 0.526.  The difference between these two values and $0.516\pm0.009$ is of marginal statistical significance, reveals no systematic trend, and is small compared to the uncertainty in the photometry.  We therefore adopt $0.516\pm0.009$ as the V-band disk fraction at the time of the SMARTS photometry, corresponding to phase 0.554.  If all disk light disappeared, the system would be 48.4\% as bright, and V-band magnitude would increase by $\Delta V=0.787\pm 0.02$.  The zero-disk V-band magnitude of the secondary at phase 0.554 is thus $(17.83\pm0.03)+(0.787\pm0.02)=18.62\pm0.04$.  

Comparing the zero-disk magnitude of the secondary star at phase 0.554 to observed passive-state magnitudes at this phase, we can now find the phase 0.554 disk fraction in each of the V- and W-band curves listed in Table 1.  For example, in curve V6, the passive state SMARTS  data, $V=18.150\pm0.01$ at phase 0.554.  This is $0.473\pm0.04$ mag brighter than the zero-disk stellar magnitude, implying a disk fraction of $0.35\pm0.03$ at phase 0.554.  Since we have performed a magnitude calibration and color correction to allow the direct comparison of V- and W-band data, we can similarly determine the phase 0.554 disk fraction for other V- and W-band curves.  We denote the disk fraction at this phase by $f_{0.554}$; the values of $f_{0.554}$ we derive are listed in Table 3.  The phase 0.554 W- and V-band disk fractions are consistently nonzero even in passive-state data: the smallest ever observed is $0.34\pm0.03$.  

We note that our V-band disk fractions are somewhat higher than those found by Gonzales Hernandez et al. (2004) and by Marsh, Robinson \& Wood (1994).  It is possible that these authors happened to observe A0620 on nights when the disk was particularly quiet, or that their observations coincide with orbital phases when the disk fraction is relatively low.  Gonzales Hernandez et al. (2004) determined their disk fraction from spectra coinciding with data in curve V6.  In Sections 3.3, 4 and 5, we will argue that the disk fraction in curve V6 is phase-variable, ranging from ~0.35 near phase 0.5 to ~0.25 near phase 0.0.  The latter value is within the errors of the disk fraction found near the V band by Gonzales Hernandez (2004), who provide neither precise times nor phases for their exposures.  Thus, depending on the phase distribution of their spectra, their disk fraction near the V band could be consistent with the higher value listed for V6 in Table 3.  

\subsection{H-band disk fractions}

We now determine H-band disk fractions.  We use the spectroscopic disk fraction in FRB07, together with simultaneous SMARTS photometry.  Unlike NSV08, the spectroscopy in FRB07 is not time-resolved.  Rather, they give only an average H-band disk fraction, from which we compute the average disk fraction in curve H6, the SMARTS passive data set.  To allow direct comparison to the V- and W-band disk fractions, we convert this phase-averaged disk fraction to $f_{0.554}$ in H6, then extend the result to determine $f_{0.554}$ in H7 as well.  

FRB07 find that the average H-band disk fraction on the nights of 2004 January 8-10 was $0.18\pm0.02$.  Their data cover every phase at least once, with triple coverage of phases 0.92-0.47 and double coverage of phases 0.47-0.49, 0.59-0.61, and 0.69-0.92.  The three nights of the FRB07 observations include five H-band SMARTS observations, all of which fall in the range of phases with triple coverage in the FRB07 data set.  The five SMARTS data points are on average $9\pm3\%$ brighter than data at the corresponding phases in H6.  We assume that the average disk fraction during these five SMARTS exposures was $0.18\pm0.02$, and that the disk accounts for all excess light relative to passive.  Under these assumptions, the stellar fraction in curve H6 is $9\pm3\%$ greater than the stellar fraction in the FRB07 data.  This implies that the average disk fraction in H6 is $0.11\pm0.03$.  

To allow direct comparison to V- and W-band disk fractions, we now convert the phase averaged H6 disk fraction to $f_{0.554}$.  To relate a phase-averaged disk fraction to the disk fraction at a specific phase, we need to determine the variation of the disk as a function of phase.  We do this by adopting a pure ellipsoidal model and attributing all deviations from the model to variation in the disk.   This procedure potentially depends on the inclination of the ellipsoidal model, so to test for inclination dependance we use models corresponding to two different values of $i$, spanning the range allowed by preliminary fits to the V- and W-band curves using the disk fractions of Section 3.1.  Specifically, fits to all V- and W-band curves give $46^\circ<i<54^\circ$, consistent with the final determination of $i$ we will make in Section 4.  We find that within this range, the choice of $i$ does not make a statistically significant difference in the value of $f_{0.554}$: adopting $i=46^\circ$ gives $f_{0.554}=0.132\pm0.03$, while $i=54^\circ$ gives $f_{0.554}=0.124\pm0.03$.  We thus adopt $f_{0.554}=0.13\pm0.03$ for light curve H6. 

Having determined $f_{0.554}$ for H6, we now determine $f_{0.554}$ for H7, the other H-band light curve in Table 3.  We use the same process as we did for the V- and W-band light curves.  Specifically, the mean magnitude of H6 at phase 0.554 is $H=14.92\pm0.02$.  Together with $f_{0.554}=0.13\pm0.03$, this implies a zero-disk magnitude of $H=15.07\pm0.04$.  For H7, we then determine $f_{0.554}$ by assuming that all additional light is nonstellar.  The results are included in Table 3.  The errors quoted include only the error inherited from the determination of $f_{0.554}$ in H6; they exclude the uncertainty in the magnitude calibration of H7 with respect to H6.  As discussed in Section 2.1, this relative calibration is particularly uncertain, so the H7 disk fraction is potentially subject to a large systematic error.  The H-band disk fractions are smaller than the V- and W-band disk fractions, but they are still significantly nonzero and therefore necessary to a robust determination of $i$.  

\subsection{The Disk Origin of Nonellipsoidal Variations}

In this section, we will compare the color of nonellipsoidal variations in the SMARTS light curves, to the color of disk light implied by the disk fractions of \S3.1 and \S3.2.  We will find that the nonellipsoidal variations have a color consistent with a disk origin.  In addition, the results of this section provide a consistency check on the results of \S3.1 and \S3.2: if our determinations of V- and H-band disk fractions were off, the implied color would likely not agree with the observed deviations from an ellipsoidal light curve. 

The disk fractions derived in Sections 3.1 and 3.2 give the V-H color of disk light, relative to star light, at phase 0.554 in the SMARTS passive light curve.  We will express this color in terms of the ratio ${{\Delta H}\over {\Delta V}}$, where $\Delta V$ and $\Delta H$ represent the difference between the star-plus-disk magnitude and the star-only magnitude.  In Section 3.1, we found $f_{0.554}=0.35\pm0.03$ in the SMARTS passive V-band curve, implying $\Delta V=0.47\pm0.04$; similarly, the SMARTS passive H-band disk fraction gives $\Delta H=.15\pm.036$.  Together, these two values give ${{\Delta H}\over {\Delta V}}=0.32\pm0.08$.

The SMARTS passive light curves exhibit prominent and wavelength-dependent deviations from pure ellipsoidal curves; we now express the color of these nonellipsoidal variations in terms of ${{\Delta H}\over {\Delta V}}$.  In each of 50 phase bins, we compute the differences $\Delta V$ and $\Delta H$ between the mean SMARTS passive V-band magnitude and the magnitude of a pure ellipsoidal light curve.   As in Section 3.2, we perform this calculation using pure ellipsoidal light curves for both $i=46^\circ$ and $i=54^\circ$ and find that varying $i$ within this range affects the results which follow by just a fraction of one standard deviation.  In Figure 3, we plot $\Delta V$ against $\Delta H$

The slope of the points in Figure 3 gives the color of nonellipsoidal variations in terms of ${{\Delta H}\over {\Delta V}}$.  We determine this slope using a principle components analysis (PCA), because $\chi^2$ minimization systematically underestimates slope when both variables have significant uncertainty.  PCA may also be biased if $\Delta V$ and $\Delta H$ have different errors, so we use a Monte-Carlo model to estimate both the bias and the uncertainty in ${{\Delta H}\over {\Delta V}}$.  The scatter in Figure 3 is consistent with photometric errors, which we therefore use in our Monte Carlo error determination.  We find a negligible bias and a slope of ${{\Delta H}\over {\Delta V}}=0.31\pm0.065$, giving the color of non-ellipsoidal periodic variations in the SMARTS passive light curve.  

\begin{figure} \figurenum{3}
\plotone{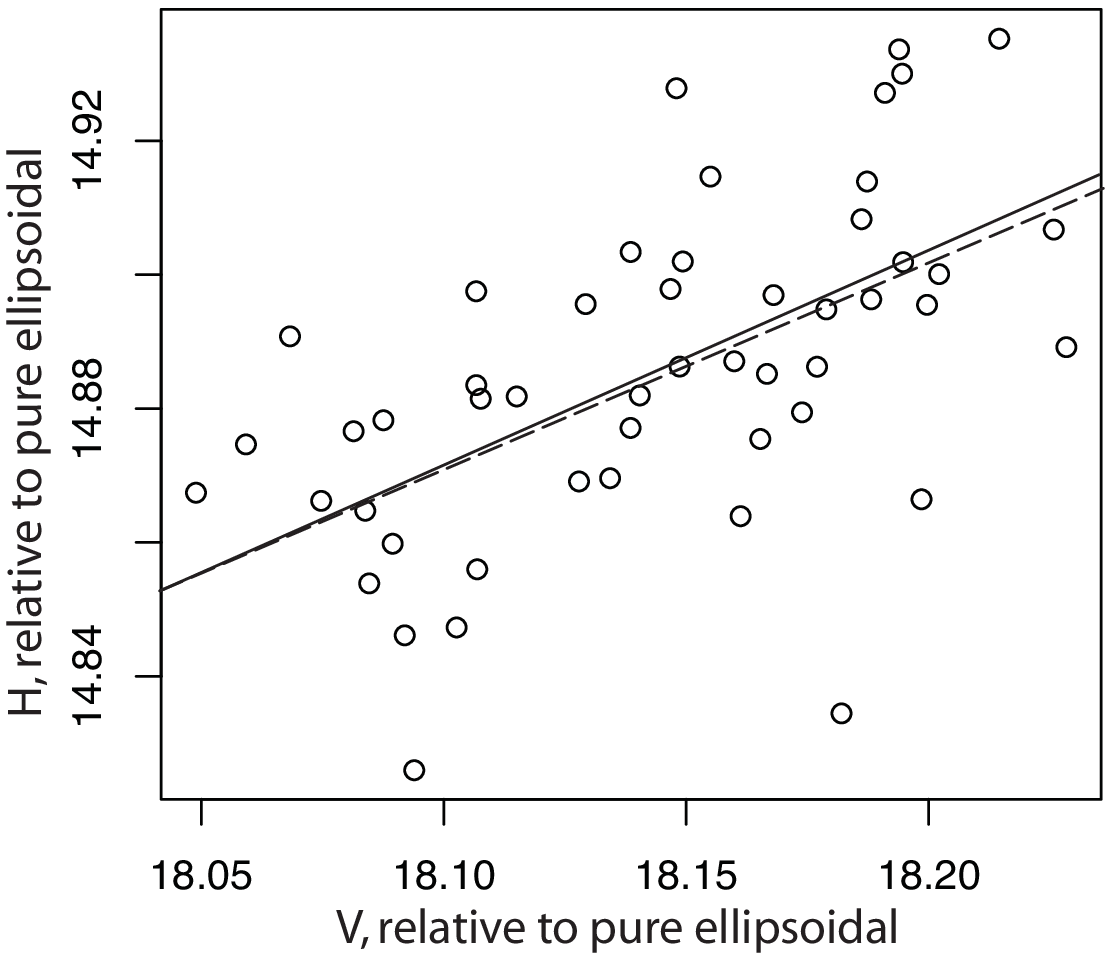}
\caption{Periodic, non-ellipsoidal variability in the V- and H-band passive SMARTS data.  In each of 50 phase bins, we compute the difference between the observed SMARTS magnitude and that of a pure ellipsoidal light curve with $i=46^\circ$ and zero mean magnitude.  If we assume $i=54^\circ$ instead, the individual points are displaced somewhat, but there is no statistically significant effect on the slope of the data.  The dashed line shows the slope implied by the points, as determined by a principle components analysis.  The solid line is not determined using the points, but rather is the slope we would expect if the non-ellipsoidal variations are due to variation a disk component with the color implied by Sections 3.1 and 3.2.  The agreement between the two lines suggests that the nonellipsodial variations originate in the disk; see text for a quantitative comparison.  The normalization of both lines is arbitrary, so we have drawn them to agree at the left edge of the plot.} 
\end{figure}

We have given two determinations of ${{\Delta H}\over {\Delta V}}$, corresponding to (1) the color of disk light identified in Sections 3.1 and 3.2 and (2)  the color of periodic, non-ellipsoidal variations. These two determinations agree to within a fraction of one standard deviation, suggesting that nonellipsoidal variations originate in the disk.   One could argue that this logic is circular, since method (1) assumed that the disk dominates all nonellipsoidal variations.  However, if this assumption is wrong, it would not have a self-consistent effect on the two determinations of  ${{\Delta H}\over {\Delta V}}$.  Specifically, method (1) depends strongly on the normalization of the light curve and only weakly on changes in shape, while method (2) depends entirely on the curve shape: in terms of Figure 3, method (2) uses only the slope of the points, not their normalization.  If stellar features produced a nonellipsoidal change in stellar flux, method (2) would merely give the color of these stellar features at the time of the SMARTS light curve.  Method (1), by contrast, would give a result depending on the spectroscopic disk fractions and the amount of excess stellar flux present in both the NSV08 data and the FRB07 data, relative to the SMARTS passive data; this amalgam of spectroscopic disk fraction and variable stellar flux would not have any simple physical interpretation, and would be unlikely to agree with the color of features in the SMARTS light curve.  

The results of this section also provide a check on the spectroscopic disk fractions from NSV08 and FRB07, on which all our disk fractions are based.  Hynes, Robinson, \& Bitner (2005) showed that there may be large systematic errors in derivations of disk fractions based on veiling.  For example, they show that if the star and template are mismatched by 100K, the derived disk fraction may be off by a factor of 2.  However, these effects would be unlikely to produce a self consistent error in the resulting V- and H-band disk fractions.   This is particularly true given that the NSV08 and FRB08 data sets were non-simultaneous and active: in order to give the right color, any error in the disk fractions would have to be consistent across a range of wavelengths and also be insensitive to variations in the disk between the two sets of observations. 

Finally, we use the slope in Figure 3, together with the value of $\Delta V$ from Section 3.1, to give a new derivation of $\Delta H$.  Specifically, ${\Delta H=\Delta V \times {{\Delta H}\over {\Delta V}}=(0.47\pm0.04)(0.31\pm0.065)=0.15\pm0.03}$.  This value and error are identical to those derived using the disk fraction of FRB07.   We improve our error on the H-band disk fraction by combining the two determinations, giving $f_{0.554}=0.13\pm0.02$, listed in Table 3.  In addition to indicating a disk origin of nonellipsoidal variations and providing a consistency check on our determinations of disk fractions, the results of this section have thus given us a new technique for determining disk fractions.

\subsection{I-band disk fractions}

In this section, we extend the method of Section 3.3 to determine the I-band disk fraction of the SMARTS passive light curve, then extrapolate as in Section 3.2 to determine disk fractions for all I-band light curves in Table 3.  This is a key determination, as half of the light curves in Table 3 are I-band data.  Following Section 3.3, we divide the SMARTS passive data into 50 phase bins and, in each bin, we define $\Delta V$ and $\Delta I$ to be the difference between the SMARTS passive data and a pure ellipsoidal light curve with $i=46^\circ$; as in Sections 3.2 and 3.3, using $i=54^\circ$ instead affects the results by just a fraction of one standard deviation.  

\begin{figure} \figurenum{4}
\plotone{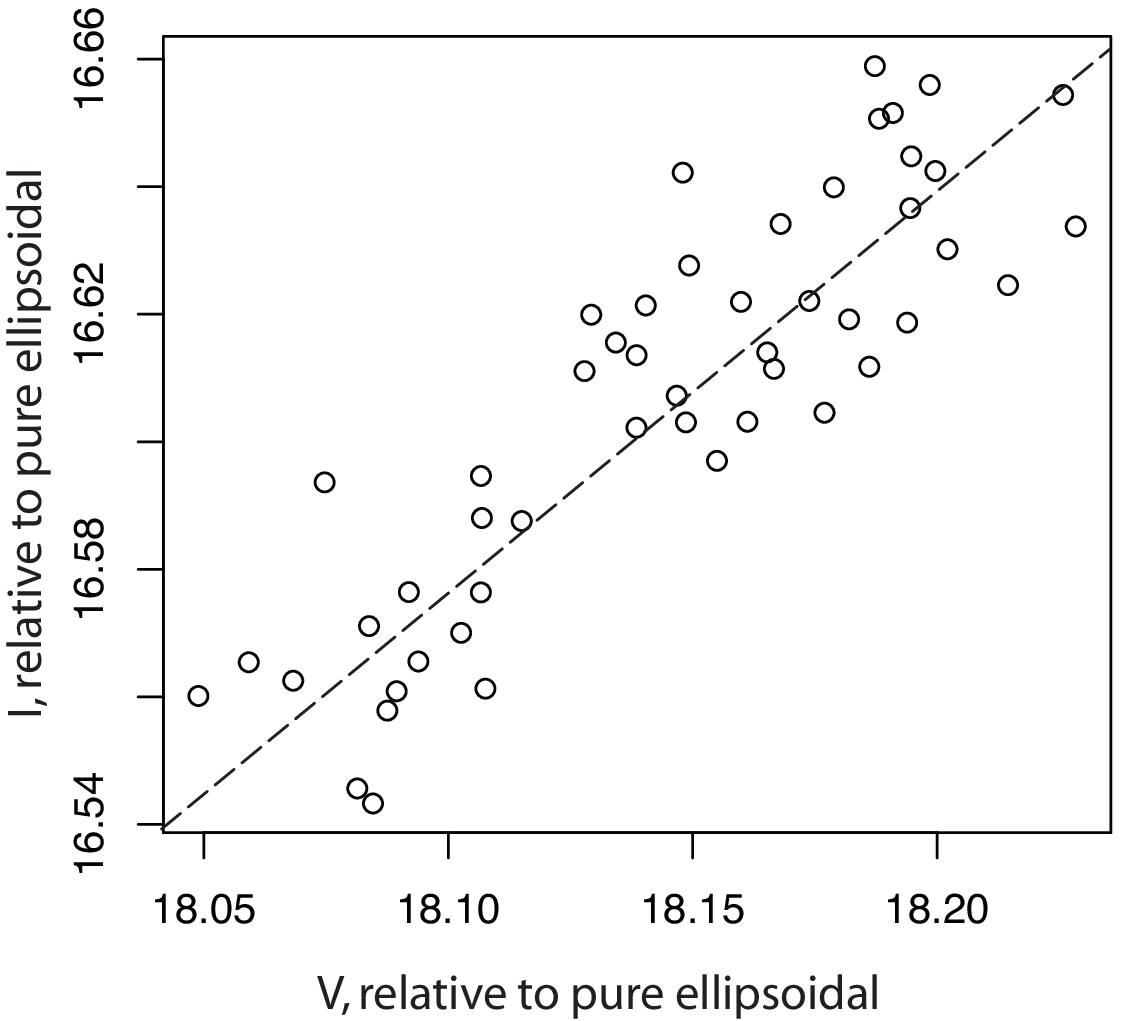}
\caption{Periodic, non-ellipsoidal variability in V- and I-band passive SMARTS data.  As in Figure 3, the points plotted come from assuming $i=46$, but  the slope is not significantly changed if we assume $i=54^\circ$ instead.  The linear relationship indicates that the nonstellar light has a consistent color throughout the orbit. The dashed line shows the slope of the points, representing the color of disk light used in our determinations of I-band disk fractions. } 
\end{figure}

In Figure 4, we plot $\Delta I$ as a function of $\Delta V$, revealing a linear relationship between the two.  As with Figure 3, we find that the scatter about the line in Figure 4 is consistent with photometric errors.  Following Section 3.3, we use a principle components analysis to determine the slope and a Monte-Carlo model to determine the error, giving ${{\Delta I}\over {\Delta V}}=(0.63\pm 0.055)$.  Together with $\Delta V=0.47\pm0.04$, determined in Section 3.1, we find that $\Delta I=(0.47\pm0.04)\times(0.63\pm0.055)=0.31\pm0.04$ at phase 0.554 in I6, the SMARTS passive light curve.  This implies $f_{0.554}=0.25\pm0.03$ in curve I6.

As in Section 3.2, we now use the SMARTS passive disk fraction to extrapolate disk fractions in other curves, attributing to the disk all photometric variability at a given phase.  At phase 0.554, the SMARTS passive curve has $I=16.62\pm0.01$ and $\Delta I=(0.47\pm0.04)$.  This implies a zero-disk I-band magnitude of $16.93\pm0.04$ at phase 0.554.  Comparing this to the observed phase 0.554 magnitude of each I-band light curve in Table 3, we compute $f_{0.554}$ for each of these curves.  The results are listed in Table 3, along with the results of Sections 3.1 and 3.2.  

The I6 and I8 disk fractions appear to be quite different, but this is an artifact of the phase we have chosen for measuring the disk fraction.  Both I6 and I8 have one very deep minimum due to the phase-variable disk, but the phasing of this minimum is opposite in the two cases: Phase 0.554 is near maximum disk brightness in I6 and near minimum disk brightness in I8.  If we had instead determined disk fractions at phase 0.0, the relative values of the disk fraction would be reversed.

\section{Determination of $i$ by fitting ellipsoidal light curves}

We now determine $i$ by using ELC (Orosz \& Hauschildt 2000) to fit ellipsoidal models.  ELC generates models of ellipsoidal light curves, with possible contributions from the disk, star spots, and X-ray heating of the secondary.  For a given star and disk configuration, it computes the light curve which would be observed through a filter with a user-specified transmission curve.  It has several optimizer routines to search the wide parameter space of possible models, finding optimal fits to data based on $\chi^2$ minimization.  In this section, we use ELC to fit models to the data listed in Table $2$ and thereby determine $i$.  

We have found that all our light curves have a nonzero disk fraction, so we include a disk in all our fits.  We constrain $f_{0.554}$ in each curve to agree with the value given in Table 2.  In addition, the results of Section 3.3 indicate that disk light dominates the non-ellipsoidal component of variation; to allow for this phase-variable disk component, we allow for a spotted disk in all models.  We fit each curve individually, rather than performing simultaneous fits on simultaneous curves.  This provides a stronger consistency check on $i$: while it is good to find a single value of $i$ consistent with all curves, it is far more compelling to find that multiple curves independently indicate the same value of $i$.  In Section 4.1, we describe the parameters involved in our models, and in Section 4.2 we make a final determination of $i$ based on the results from individual curves.

\subsection{Input parameters used in determining $i$}

The ELC model as applied to A0620 has many parameters, several of which can be set to reasonable values based on photometric and spectroscopic observations of the system.  The most important parameters are related either to the basic geometrical properties of the binary and the components therein or to the radiative
properties of the star and accretion disk.

The parameters that set the scale of the binary and determine the shape of the star include the orbital period $P$, the orbital separation $a$, the ratio of the masses $Q\equiv M/M_2$, where $M$ is the mass of the black hole and $M_2$ is the mass of the secondary star, the inclination $i$, and the Roche lobe filling factor $f_2$.  In the case of A0620, the orbital period $P$ is quite well known  and was held fixed at $P=0.3230160$ d (Johannsen et al.\ 2009).  We assume the star fills its Roche lobe, and consequently we set $f_2=1$.  We also assume the star rotates synchronously and that the orbit is circular.  Finally, the search of parameter space is more efficient if we use the mass of the secondary star $M_2$ (a free parameter) and the semiamplitude of its radial velocity curve $K_2$ (taken from NSV08) to set the scale of the binary. 

The parameters related to the radiative properties of the star include its average temperature $T_2$, its gravity darkening exponent $\beta$, and its bolometric albedo $A$.  The value of $T_2$ was set to 4600 K (GHO01), and the gravity darkening exponent was set to 0.1 (Claret 2000).  There is some disagreement on the temperature of A0620 (e.g. FRB07 find T=4000-44000), but this small ambiguity in T will not change out results significantly, so we use only the GHO01 value.  Since A0620 is exceedingly faint in X-rays in quiescence (Garcia et al. 2001), the effects of X-ray heating were neglected.  Since ELC uses specific intensities derived from model atmosphere computations, no parameterized limb darkening law is needed.  The specific intensities used for A0620 were derived from the {\sc NextGen} grid (Hauschildt, et al.\ 1999a, 1999b) with updates (Hauschildt private communication).

Since all curves in Table 2 have a nonzero disk fraction, all our fits include a disk.   The disk in ELC is circular and flared, and the parameters describing it include the outer radius of the disk $r_{\rm out}$, the inner radius of the disk $r_{\rm in}$, which was fixed at 0.001 times the black hole's Roche lobe radius, the opening angle of the outer disk rim $\beta_{\rm rim}$, the temperature of the inner edge $T_{\rm disk}$, and the power-law exponent on the disk temperature profile $\xi$, where $T(r)=T_{\rm disk}(r/r_{\rm in})^{\xi}$.  Since optical emission is dominated by the outer disk, the precise value of the inner radius is unimportant to the optical emission discussed here.

Our models all include a hotspot on the disk, since we have argued in Section 3.3 that disk light --- not star light --- is the dominant source of periodic, non-ellipsoidal variability.  In all cases, we find that one spot is sufficient to fit the data.  In the ELC model, the spot is on the rim and extends down to a radius $r_{\rm cut}$ on the disk face.  The temperature on each pixel within the spot is some constant $s_{\rm spot}$ times the underlying temperature.  The spot is centered at azimuth $\theta_{\rm spot}$, and has a radius $w_{\rm spot}$ (in degrees).  Thus one needs four additional parameters to model a spot on the disk.

We note that at low inclinations, the spot is generally in view for most or all of the orbit: it is not visible at some phases, then hidden at others.  Rather, the spot sits on the surface of a flared disk and changes in projected intensity as our viewing angle of the flared disk changes.  This results in a smooth variation in brightness throughout the orbit, in contrast to the sharp dip which would be present if the spot were ever eclipsed, by either the disk or the star.  In addition, the spot color is consistent throughout the orbit since the variation in flux is due to a projection effect, not the appearance and disappearance of extra-hot regions.  These characteristics --- smooth variation and constant spot color throughout the orbit --- are apparent in Figures 3 and 4, where an eclipsed feature would appear as a discontinuous jump in color and magnitude.  

To summarize, we have 11 free parameters:  $i$, $M_2$, $K_2$, $T_{\rm disk}$, $r_{\rm out}$, $\beta_{\rm rim}$, $\xi$, $r_{\rm cut}$, $s_{\rm spot}$, $\theta_{\rm spot}$, $w_{\rm spot}$.  We have an additional three extra constraints on the model.  NSV08 measured a $K$-velocity of $K_2=435.4 \pm 0.5$ km s$^{-1}$, and a projected rotational velocity of $V_{\rm rot}\sin i=82\pm 2$ km s$^{-1}$.  These two constraints are imposed by applying a penalty to $\chi^2$, equal to the weight of any one data point.  Finally, in each light curve, we likewise constrain the disk fraction $f_{0.554}$ by applying a penalty to $\chi^2$ for deviations from the value given in Table 2.  

The fits were optimized by using three of ELC's available optimizers: a scheme that uses a genetic algorithm, an algorithm using a Monte Carlo Markov Chain, and a scheme using a downhill simplex method.  These optimizers were used in combination, with the results from each contributing to the final $\chi^2$ curve.  We fit each curve individually, even in data sets which were obtained simultaneously.  The inclinations given by our fits are listed in Table 3; the best fits and residuals are shown in Figure 2.  

\subsection{Determination of $i$}

In this section, we give a determination of $i$ with the smallest uncertainty that can be achieved based on the twelve measurements of $i$ listed in Table 3.  The values of $i$ in Table 3 are generally susceptible to four sources of error: (1) any mismatch between the SMARTS magnitude calibration and the magnitude calibration in other data sets, leading to an error in $f_{0.554}$ and thus in the derived value of $i$; (2) Systematic errors in our fitting procedure, e.g. imperfect modelling of the disk or star, or problems in our method of determining disk fractions; (3) the statistical errors quoted in Table 3; and (4) Systematic errors due to the fact that all disk fractions are derived from just two spectroscopic observations and are therefore not independent.  We will consider each of these errors in turn.

We begin by considering the error due to magnitude calibration relative to the SMARTS data set.  Since we are concerned only with {\it relative} calibration, the SMARTS data set is not susceptible to this error; data sets with calibration problems will therefore give determinations of $i$ which differ systematically from the values given by SMARTS curves.  Taking a weighted mean of the values of $i$ given by the SMARTS passive light curves gives $i=51\fdg13\pm 0.85$.   Of the nine non-SMARTS curves in Table 3, six  curves (W1, I1, W2, W3, I4, and I8) give values of $i$ consistent with the SMARTS value: among these six curves, the biggest deviation from the SMARTS value is $1.54\sigma$, consistent with what one would expect from statistical uncertainty alone.  Given that there is little room for any non-statistical source of disagreement among these curves, calibration errors, if any, must be small.  

The three curves which deviate significantly from the SMARTS value of $i$ all come from data sets identified in Section 2.1 as being particularly susceptible to magnitude calibration errors: (1) H7 comes from FR01, in which differential magnitudes of various comparison stars are inconsistent with the SMARTS values, and (2) I2 and I3 come from the McGraw-Hill I-band data set, in which the I-band data only had one comparison star and we therefore could not perform a color correction.  Moreover, I1, I2 and I3 indicate a self-consistent calibration error within this data set: I1, which agrees with the SMARTS value of $i$, is the curve whose fit is least sensitive to the disk fraction and thus to the magnitude calibration.  Specifically, if the magnitude calibration is off by ~0.12 mag, all three curves would give inclinations within $1.5\sigma$ of the SMARTS value.  A consistent calibration error thus puts these curve in better agreement with each other and with the SMARTS curves.  We conclude that magnitude calibrations and disk fractions in I1, I2, and I3 and H7 are likely off, and we therefore exclude these curves from subsequent analysis of $i$.

We now consider the second source of error mentioned above, namely the possibility that our fitting procedure gives spurious values of $i$ due to either imperfect modeling by ELC or faulty assumptions in our method of determining disk fractions.  We have seen above how spurious disk fractions lead to conspicuous errors in $i$, so the agreement between V6, I6, and H6 supports our determinations of the disk fractions.  If, for example, we used the disk fractions given by either GHO01 or Gonzales Hernandez et al. (2004), V6, I6 and H6 would give inconsistent results.  Moreover, a self-consistent error in these disk fractions seems improbable: if the assumption of constant stellar flux at a given phase were wrong, we would expect V6 and H6 to give inconsistent results, given that the two spectra are from different epochs.  Similarly, if the color of nonellipsoidal variations in these curves did not indicate the color of the disk, then the I-band fraction would be off and we would expect I6 to give a different inclination than V6 and H6.  Our assumptions are further supported by the agreement between V6 and the W-band curves, the agreement between I6, I8, and I4, and the general agreement between W-band and I-band data.  

The agreement between different curves also suggests that any failures of the models used by ELC do not result in a significant error in $i$.  Most obviously, if ELC's simple disk model was flawed in a way which yielded a spurious ellipsoidal component, we would expect the curves with strong nonellipsoidal features --- V6, I6, H6, and I8 --- to give inclinations different than curves with minor disk features ---  W1, W2, W3, and I4.  In addition, the agreement among the SMARTS curves indicates that the disk model produces no systematic error in $i$: the disk is more prominent in bluer bands, so if  we were over- or under-compensating for the disk, we would expect a systematic trend in $i$ between the different bands; no such trend is present.  In addition, the agreement between curves in different bands supports the robustness of the pure ellipsoidal light curves generated by ELC.  For example, if the code used a nonphysical model of limb darkening, it could produce a systematic error in the amplitude of light curve associated with a given $i$.  Such an error, however, would probably not be consistent across several bands.  The agreement between results from curves in different bands, with different nonellipsoidal components, therefore supports the robustness of the models used in ELC.  

We now consider the statistical error, which we reduce by combining together the results from many curves.  Reducing the statistical error in this manner is worthwhile only if the systematics are small, but we have argued that they must be: while simultaneous curves share some systematics, curves in the same band share others, and curves with a relatively large disk component share yet others.  The agreement across these various groups indicates that the corresponding systematic systematic effects are all small.  The statistical errors are in good agreement with the scatter present in our values of $i$, indicating that the formal errors given by fitting are realistic.  For example, the inclinations given by V6, I6, and I8 all have errors below $1.5^\circ$ and give values of $i$ ranging from 50.13 to 51.75. The other five light curves with reliable magnitude calibrations have larger errors, but taking a weighted mean of their results gives  $i=50\fdg93\pm1.5$.  We therefore have four statistically independent measurements of $i$, each with errors near $1.5^\circ$, which span a range of just 1.62 degrees, smaller than would be expected given the statistical errors.  This indicates that the formal errors given by the fits are realistic and, if anything, are overestimated.  We therefore compute a refined value of $i$ by taking a weighted mean of the results from W1, W2, W3, I4, V6, I6, H6, and I8.  The weighted mean of these eight measurements of $i$ is $i=50\fdg98\pm0.71$, which we adopt as our final determination of $i$ and the statistical error on $i$.  

Finally, we consider the systematic error due to the fact that all determinations of $i$ originate from just two spectroscopic measurements of the disk fraction.  In particular, all W-, V- and I-band disk fractions are based on the same spectroscopic measurement in NSV08; seven of the eight curves used to determine $i$ will therefore suffer from the same systematic error if the result from NSV08 is off.  To estimate this systematic error, we fit every V- and I-band curve with disk fractions offset by $\pm\sigma$, then recompute the weighted mean inclination, once with systematically large disk fractions and once with systematically small disk fractions.  We find that the weighted mean changes by $\pm0.5$ when all V- and I-band disk fractions are systematically offset by 1 $\sigma$, and we adopt this as the systematic error due to the common origin of the disk fractions.  Combining this in quadrature with the statistical error of 0.71, we find $i=50\fdg98\pm0.87$; this error includes both statistical error and the only systematic error which we believe is significant. 

As a final check on the value of $i$, we fit an ellipsoidal light curve with a spotted disk to the H-band light curve in GHO01.  This curve was not included in Tables 1 and 2, because we were able to obtain the data only in binned form and therefore cannot be sure that it is passive and consistent in shape throughout the observations.  We therefore find these results less trustworthy than others, but view them as a useful check on our determination of $i$.  Following the method of Section 3.2, we find a disk fraction of $0.2\pm0.05$.  Assuming this disk fraction and fitting the curve with a model including a spotted disk, we find $i=48^\circ\pm3$, consistent with the value of $i=50\fdg98\pm0.87$ determined from the curves in Table 3.  The lower value of $i$ determined by GHO01 is therefore completely explained by their decision to exclude the disk.

In conclusion, we find the inclination of A0620 to be $i=50\fdg98\pm0.87$, implying a black hole mass of $M=6.61\pm0.25M_\odot$ (NSV08) and a stellar mass of $M_2=0.40\pm0.045M_\odot$.  Of the twelve curves in Table 3, nine are consistent with this value.  The only deviants --- I2, I3, and H7 --- are likely off due to  problems with their magnitude calibration.  The curves with reliable magnitude calibrations have a spread consistent with that expected from statistical errors alone, indicating that most sources of systematic error are small; in particular, the assumption of constant stellar flux at a given phase, and the modeling of the star and disk in ELC, appear to be robust against systematic errors in $i$.  The data sets included in this paper previously gave inclinations spanning nearly 40 degrees, but we have brought them into agreement by excluding active state data, combining light curves only if they show a consistent shape, and accounting for disk light on a curve-by-curve basis.

\section{The Distance to A0620-00}

In this section, we follow the method of  Barret, McClintock \& Grindlay (1996) to estimate the distance to A0620: we compute the absolute magnitude implied by the spectral type and radius of the secondary, then use its apparent magnitude and reddening to estimate the distance.  We use Eggleton (1983) to compute $R_2$, the effective radius of the Roche lobe.  We then use Popper (1980) to compute the absolute magnitude based on $R_2$ and the spectral type.  Finally, we use the reddening law of Fitzpatrick (1999) to compute dereddened apparent magnitudes and compute the distance.  We note that Popper (1980) assumed $\log(g/g_\odot)=0.1$ for a K5V star, while our dynamical model implies  $\log(g/g_\odot)=-0.05\pm0.05$; we will argue later that A0620's low surface gravity is  of minimal importance to the distance determination, and therefore use Popper (1980) and other works without an adjustment for surface gravity.  Our robust determinations of $i$ and multi-band disk fractions eliminate two of the major uncertainties in the distance determination, namely the radius of the secondary and the flux contribution of the disk (Barret, McClintock \& Grindlay 1996).  

Two major uncertainties remain in the computation of distance: both the reddening and the spectral type of A0620 have been disputed, and a range of values for each has been quoted in the literature.  We will show that these uncertainties have a relatively small effect on distance when constrained by the zero-disk color of A0620, but we begin by considering the full range of both values which have been claimed in the literature.  The reddening of A0620 has been determined by many methods: use of the 2175\AA\ feature, Na D lines, dust maps, and comprehensive modeling of diffuse interstellar bands.  These various methods give values of $E(B-V)$ ranging from 0.25 to 0.49.   Hynes (2005, and references therein) argues that the most robust value is $E(B-V)=0.35\pm0.02$ (Wu et al. 1983), and that the only measurement less than 0.3 is strongly model-dependant.  

The most reliable determinations of spectral type come from NIR spectroscopy obtained by FRB07 and Harrison et al. (2007), who find spectral types of K5 and K5-7, respectively.  These spectral types are in good agreement with the pioneering work of Oke (1977), who also found a spectral type of K5-7.  Spectral types of K4 (GHO01) and K2-3 (Gonzales Hernandez et al. 2004) have also been claimed, but there are reasons to be suspicious of both results.  Specifically, GHO01 used only photometric data and neglected the disk contribution, while Gonzales Hernandez et al. (2004) determined the spectral types and disk fraction based exclusively on the FeI lines;  FRB07 argued that this may have biased their results.  In addition, the results of Gonzales Hernandez et al. (2004) imply that the disk is negligible redward of 7500\AA\ , contrary to the significant I- and H-band disk fractions we have found.  A spectral type of K5 is thus preferred by existing observations, but for the sake of completeness we will consider spectral types of K2, K4, K5, and K7, spanning the full range of values which have been claimed in the literature.

Though both the reddening and the effective temperature are uncertain, these two uncertainties are not independent: given the zero-disk colors derived in Section 3, a hotter star requires more reddening in order to match the observed color.  Since the color of the secondary varies as a function of phase, we compute $E(B-V)$ at both phase 0.5 (a minimum) and phase 0.25 (a maximum).  In Section 3, we computed zero-disk magnitudes at phase 0.554 only.  We therefore use ELC to model the variation in VIH stellar magnitudes as a function of phase, normalizing the model VIH stellar magnitudes to agree with our zero-disk magnitudes at phase 0.554.  For a given spectral type and phase, we assume that the star has the intrinsic colors given by Bessell \& Brett (1988), assume the reddening law of Fitzpatrick (1999), and compute the value of E(B-V) implied by the apparent zero-disk V-H color.   To check for consistency, we use the same procedure to compute the reddening implied by the observed V-I and I-H zero-disk colors.  Where these determinations of E(B-V) are inconsistent, it indicates either a nonstandard reddening law or a mismatch between the assumed spectral type and A0620.  Table 4 lists the values of $E(B-V)$ derived from various spectral types and phases, based on the apparent colors of A0620.  We use subscripts to denote the color used in each determination of $E(B-V)$.  Since  $E(B-V)_{V-H}$ has the smallest statistical error, we use this value to compute the dereddened magnitude $V_0$ and distance, again assuming the reddening law of Fitzpatrick (1999).  Our errors on $E(B-V)_{V-H}$, $V_0$, and $d$ all include any disagreement between $E(B-V)_{V-I}$ and $E(B-V)_{I-H}$.

\begin{deluxetable}{llllllll}
\tablecaption{Distance and Reddening of A0620}
\tabletypesize{\footnotesize}
\tablehead{
\colhead{MK Type} &\colhead{Phase} & \colhead{$E(B-V)_{V-I}$\tablenotemark{a}} & \colhead{$E(B-V)_{I-H}$\tablenotemark{a}} & \colhead{$E(B-V)_{V-H}$\tablenotemark{b}} & \colhead{$V_0$\tablenotemark{c}} & \colhead{$M_V$\tablenotemark{d}} & \colhead{$d$(kpc)}}

\startdata
K2V &0.5&   $0.45\pm0.04$ & $0.62\pm0.05$ & $0.52\pm0.085$ & $17.13\pm0.27$ & $6.61\pm0.05$ & $1.26\pm0.35$\\
K2V &0.25& $0.42\pm0.04$ & $0.59\pm0.05$ & $0.49\pm0.085$ & $16.98\pm0.27$ & $6.61\pm0.05$ & $1.17\pm0.33$\\
K4V &0.5&   $0.34\pm0.04$ & $0.42\pm0.05$ & $0.38\pm0.04$ & $17.47\pm0.13$ & $7.20\pm0.05$ & $1.13\pm0.16$\\
K4V &0.25& $0.31\pm0.04$ & $0.39\pm0.05$ & $0.34\pm0.04$ & $17.32\pm0.13$ & $7.20\pm0.05$ & $1.05\pm0.14$\\
K5V &0.5&   $0.30\pm0.04$ & $0.30\pm0.05$ & $0.30\pm0.02$ & $17.71\pm0.07$ & $7.49\pm0.05$ & $1.10\pm0.09$\\
K5V &0.25& $0.27\pm0.04$ & $0.27\pm0.05$ & $0.27\pm0.02$ & $17.55\pm0.07$ & $7.49\pm0.05$ & $1.02\pm0.08$\\
K7V &0.5&   $0.16\pm0.04$ & $0.25\pm0.05$ & $0.19\pm0.04$ & $18.05\pm0.13$ & $8.05\pm0.05$ & $1.00\pm0.14$\\
K7V &0.25& $0.13\pm0.04$ & $0.21\pm0.05$ & $0.16\pm0.04$ & $17.89\pm0.13$ & $8.05\pm0.05$ & $0.92\pm0.13$\\
\enddata

\tablenotetext{a}{$E(B-V)_{V-I}$ is the extinction derived from the observed zero-disk V-I color at phase 0.5, while $E(B-V)_{I-H}$ is derived from the zero-disk I-H color at phase 0.5.  In both cases, we assume for each spectral type that the star has the intrinsic colors given in Bessell \& Brett (1988) and we use the reddening law of Fitzpatrick(1999) to derive the value of $E(B-V)$ implied by the observed zero-disk colors.  The errors quoted are statistical errors due to uncertainty in our photometric calibrations.  Inconsistencies between the two derivations of $E(B-V)$ indicate either a nonstandard reddening law or a poor fit to the true spectral type.}

\tablenotetext{b}{$E(B-V)_{V-H}$ is the extinction derived from the observed Zero-disk V-H color by the same procedure as $E(B-V)_{V-I}$ and $E(B-V)_{I-H}$.  Since this value has the smallest statistical uncertainty and reflects color over the full color range of our observations, we use these values for the computation of $V_0$.  The quoted errors are the larger of (1) the statistical error due to photometric uncertainty, and (2) half the range spanned by $E(B-V)_{V-I}$ and $E(B-V)_{I-H}$.  The latter error reflects uncertainty in $E(B-V)$ due to inconsistencies between different bands; this is the dominant source of uncertainty for all spectral types except K5.}

\tablenotetext{c}{Dereddened, zero-disk V-band magnitude of A0620 at phase 0.5.  Computed using the observed zero-disk brightness at this phase and $E(B-V)_{V-H}$ as given in column 4.  Errors on $V_0$ are propagated using the errors listed in column 4} 

\tablenotetext{d}{Absolute visual magnitude computed assuming the star has radius $R_2$ and using the relation of Popper(1980).  Popper(1980) does not give a value of $F_V'$ for spectral type K4, so this value was computed by linear interpolation between K2 and K5}

\end{deluxetable}

Table 4 shows that K5 is the only spectral type for which the V-I color and I-H color give consistent values of $E(B-V)$.  In addition, K4-5 are the only spectral types which imply values of $E(B-V)$ consistent with any previous determination of $E(B-V$.  The spectral type of K2 is strongly inconsistent with our data, implying inconsistent and implausibly high values of $E(B-V)$.   For spectral types of K7 and K4, $E(B-V)_{V-I}$ and $E(B-V)_{I-H}$are inconsistent only at the $~1.3\sigma$ level, which could plausibly be due to a nonstandard reddening law.  However, assuming a spectral type of K7 implies an implausibly low value of $E(B-V)$.   Though K4 cannot be strongly ruled out, K5 gives the best match to the observed colors for a standard reddening law, and is also the value preferred by the most robust spectroscopic studies (e.g. FRB07). 

Table 4 shows that the distance is only a weak function of the spectral type assumed.  Assuming a hotter star makes the star more luminous, but requires greater reddening; these two changes have opposite effects on the derived distance.  For spectral types K2-7, the distance determined from phase 0.5 ranges from 1.00-1.26kpc; without the change in implied reddening, this range of spectral types would give nearly a factor of 2 uncertainty in the distance.  Thus,  our constraint on the color eliminates over 70\% of the uncertainty inherited from the uncertain spectral type.  Similarly, the precise value of $E(B-V)$ has a relatively small effect on the derived distance: given the observed zero-disk colors, larger values of $E(B-V)$ imply hotter stars, counteracting the effect of the reddening on distance.  Indeed, if we had computed spectral type as a function of $E(B-V)$ for the full range of $E(B-V)$ that has been claimed in the literature, we would actually have found a smaller range of distances than that listed in Table 4.

We note that the determination of distance is also only weakly dependent on the assumed value of surface gravity.  As with changing the assumed spectral type, decreasing $\log(g/g_\odot)$ makes the assumed star both less luminous and redder; this forces a lower value of $E(B-V)$, counteracting the effect of the lower luminosity when distance is calculated.  As an extreme example, consider computing the distance as in Table 4 assuming a K5 giant star in place of a K5 dwarf: one finds that the distance differs by just 16\%; at a constant value of $E(B-V)$, the difference in distance would be 65\%.  In A0620, $\log(g/g_\odot)$ differs from a main sequence star by just 0.15, whereas assuming a K5III  changes $\log(g/g_\odot)$ by 2.8.  Given that even this large change in the assumed $\log(g/g_\odot)$ has a relatively small effect on the derived distance, we are confident that we have not introduced a significant error by assuming a normal main sequence $\log(g/g_\odot)$ in using Popper (1980) and Bessell \& Brett (1988).

The derived distance is also a relatively weak function of phase: the distance derived at the maximum is just 0.08 kpc closer than that derived at minimum, comparable to the standard deviation on the measurement based on a spectral type of K5.  Given that the precise orbital phase used to compute the distance is of relatively minor importance, we estimate the distance simply by averaging the values given by phase 0.5 and phase 0.25.  We then incorporate this range into the error on the final value.  For our preferred spectral type of K5, this gives $d=1.06\pm0.12$~kpc.  Even if the spectral type is off, the effect on the distance is not likely to be large: spectral types of K2 and K7 are improbable given our data, and assuming a spectral type of K4 changes the derived distance by just $0.25\sigma$.  

\section{Discussion}

The results of this paper have broad implications for determinations of $i$ based on modeling of ellipsoidal light curves. The state changes we have found in A0620 are likely present in many or all of the other seven black hole binaries with periods less than half a day, such as Nova Mus 1991 and GS 2000+25 (Remillard \& McClintock 2006).  In addition, the presence of a significant and variable disk component is likely to be a feature of light curves in many other systems.  A broadband SED is not sufficient to detect the disk: though the infrared colors of A0620 are consistent with purely stellar light (GHO01), the disk is significant in these bands.  We therefore believe that future determinations of $i$ must include identifications of states, attention to possible changes in the light curve, spectroscopic determinations of disk fraction and models allowing for a phase-variable disk component.  

In Section 3.3, we showed that the color of nonellipsoidal variability in the SMARTS data is consistent with the disk color, and we concluded that the disk dominates nonellipsoidal variations.  This conclusion is supported by the success of a phase-variable disk in explaining a wide range of light curves with a consistent value of $i$ (Section 4).  A phase-variable disk, far from merely diluting a light curve and thereby reducing its amplitude, may have fundamental effects on the shape of the light curve: in some cases it may even {\it increase} the light curve's amplitude.  To illustrate the range of these effects, Figure 5 shows the best fit models of Figure 2, broken down into star and disk components.  Included are all light curves except I1, I2, I3, and H7, which have questionable determinations of $f_{0.554}$ (see Sections 4.2 and 2.1), and therefore may not yield accurate disk models. 

\begin{figure} \figurenum{5a}
\plotone{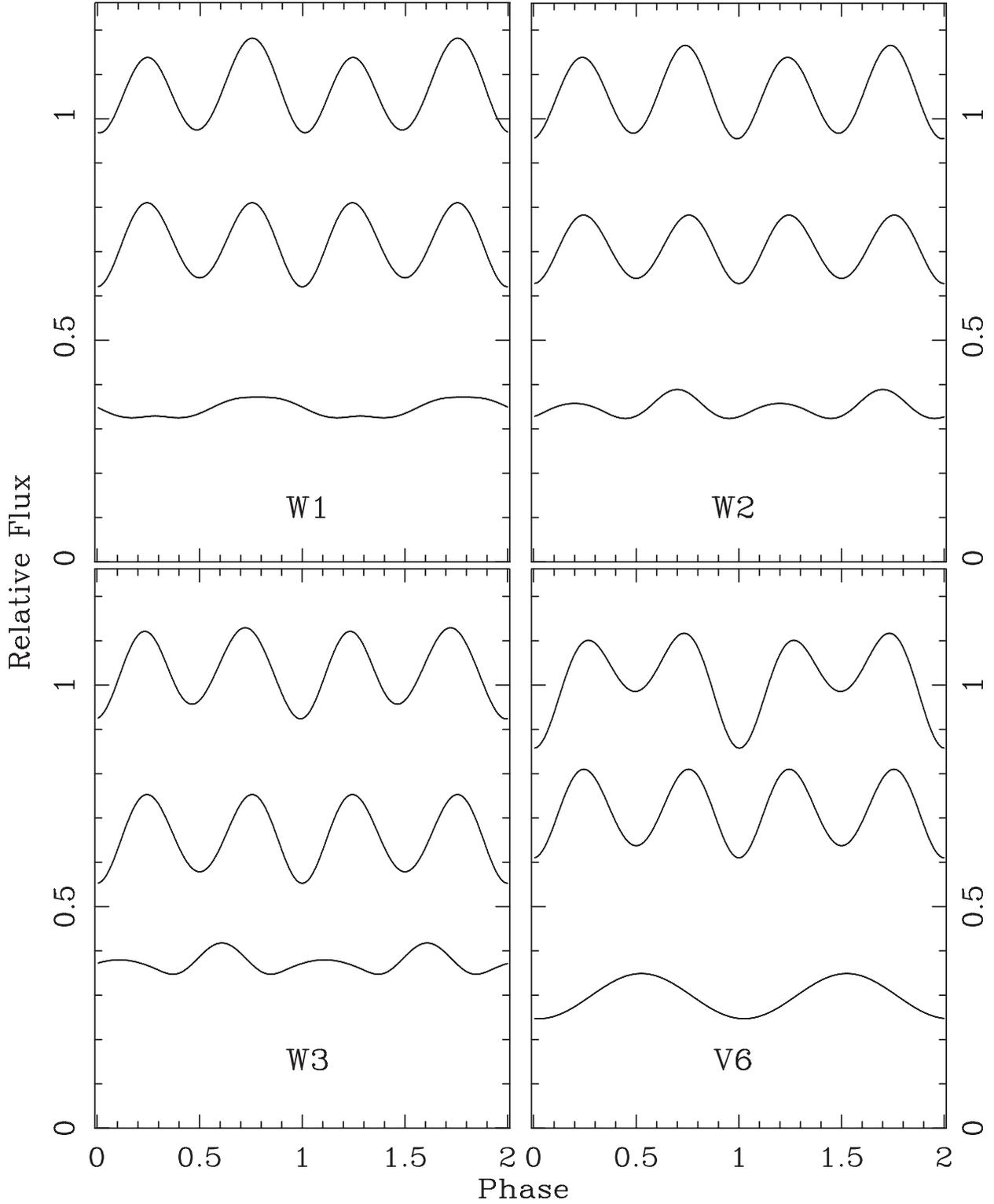}
\caption{Components of best fit ELC models for all curves with reliable values of $f_{0.554}$.  Curves are the same as the models shown in Figure 2.  In each panel, the top line shows the total light curve, the middle line shows the ellipsoidal component, and the lower line shows the phase-variable disk contribution.  All light curves are plotted in relative flux units, scaled to give total flux 1 at phase 0.554. Figure 5a: W1, W2, W3 and V6; Figure 5b: I4, I6, I8, and H6.} 
\end{figure}

\begin{figure} \figurenum{5b}
\plotone{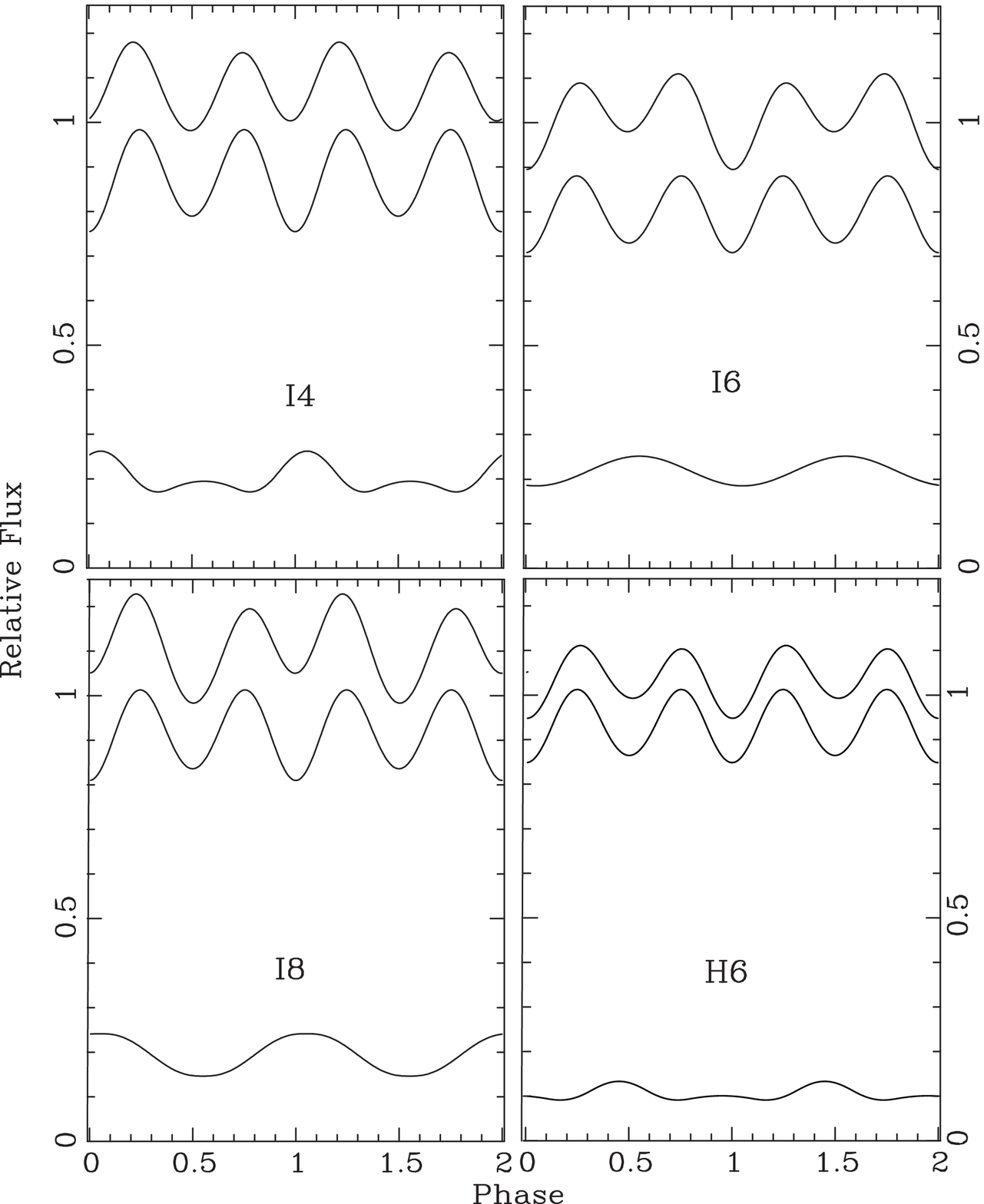}
\caption{}
\end{figure}

Figure 5 shows that even the curves which lack conspicuous nonellipsoidal features are best fit with significant variation in the disk.  There are two types of disk light curves: ones like V6, I6, and I8 which are roughly sinusoidal, and ones like W2, W3, and I4 which are double-humped; W1 and H6 are intermediate, with a flattened minimum but only one prominent maximum.  The sinusoidal disk curves generally lead to more conspicuously nonellipsoidal light curves, not because they have a stronger effect on the shape of the light curve, but because their single-waved form produces uneven maxima and minima.  

The sinusoidal and double-waved disk curves are easily understood in the context of ELC's disk model.  All models include a single spot on the rim, extending partway down the face of the disk; the relative importance of the face (sinusoidal) and rim (second peak) determine the type of light curve observed.  The face component of the spot is always visible, but its projected intensity changes sinusoidally with viewing angle, reaching a maximum when the spot is on the far side of the disk and therefore most directly facing the observer.  By contrast, the rim component is eclipsed when the spot is on the far side of the disk, then produces a relatively peaked maximum when the spot is on the near side of the disk.  The rim component is thus in antiphase with the face component, yielding a second maximum if the rim produces a significant fraction of the total spot emission.  In V6 and I6 the face is dominant, but in H6 the rim is significant compared to the face.  This reflects the temperature gradient in the disk: the face is relatively important in the optical, since the spot extending down the face is hotter than the rim.

We note that the double-waved disk features apparent in W2, W3, and I4 can easily appear to be part of the ellipsoidal modulation: to the eye, W3 and I4 look like purely ellipsoidal curves. Indeed, it is remarkable that ELC can pick out the double-waved disk component reliably enough to give a consistent value of $i$ across these curves.  The double-waved disk feature also gives a sense of why these curves cannot be well-modeled using a spotted star rather than a spotted disk: Though a single disk spot naturally produces a double-waved feature, a single starspot produces just a single-peaked modulation.  Thus, when a star spot is invoked, much of the double-waved disk feature is modeled as ellipsoidal variability, resulting in inconsistent determinations of $i$ and large, phase-variable residuals in most curves; this is just what we found when we attempted to model these curves with a spotted star and a constant disk. 
 
Though we have given a rough physical interpretation of the disk light curves, we believe that more detailed physical models are necessary in order to have real insight into the physics of the spot discussed here.  The actual disk and spot are presumably not the perfect geometric figures assumed by ELC, and there are degeneracies which allow different configurations to produce identical disk light curves.  Without more physical disk models, we cannot know whether the spot we have found comes from the optically thin disk which produces the spot seen in emission line studies (e.g. NSV08), or whether it is coming from an optically thick component of the sort described by Hynes, Robinson, \& Bitner (2005). Nor can we say with certainty whether this spot is related to the stream-disk impact point or is due to some other source of uneven disk heating.  It would be fascinating to model these light curves using full MHD disk simulations with radiative transfer.  In particular, we note that the data in figure 1 show that the configuration which produced I8 was apparently unstable, and changed on a timescale of days.  A study of this time variability could be a valuable tool in understanding the nature of the spot.

The disk features discussed here --- like the constraint provided by phase shifts (Cantrell \& Bailyn, 2007) --- reflect the third dimension of disk structure, which cannot be probed by Doppler tomography.  While the properties of phase shifts depend primarily on the velocity field and opening angle of the disk, the modulations discussed here depend primarily on the physical location of bright spots on a three-dimensional disk.  Modulations in disk brightness, phase shifts, and Doppler tomography therefore provide complementary constraints on the structure of disks in accreting binary stars.

We have demonstrated that a single determination of the disk fraction with simultaneous photometry may be sufficient to determine $i$, but future work would be improved by many independent determinations of disk fraction: the ideal data set for determining inclinations would be a passive light curve with simultaneous phase-resolved spectroscopy.  In some objects, extended periods of activity make it unfeasible to obtain passive data with simultaneous spectroscopy: CBMO08 and ongoing SMARTS observations indicate that A0620 has been active since 2003, and that Nova Mus 1991 has been persistently active for even longer.  It would nonetheless be possible to obtain a disk-corrected light curve from a persistently active object by obtaining photometry with perfectly simultaneous spectroscopy corresponding to every photometric exposure.  Then each photometric measurement could be corrected for the disk contribution present {\it in that exposure}, yielding a disk-free light curve which should be well-fit by a pure ellipsoidal model. 

It would be tempting to simply bin active state data or use some other smoothing algorithm to produce a light curve for fitting, but such procedures have fundamental problems, beyond the challenges intrinsic to working with lower-quality data: (1) ELC and similar codes assume that the disk is not intrinsically variable, with all variation due to changing viewing angle.  The active disk, however, varies rapidly in intrinsic brightness (NSV08), and smoothing over such fluctuations will not generally produce a curve which corresponds to an intrinsically constant disk.  (2) Active state data generally have upper and lower envelopes with distinct shapes.  For example, in LHO98 Figures 2b and 2c, there is excess variability near phase 0.5, so the depth of this minimum changes by ~0.1 magnitude depending on whether one considers the mean, the upper envelope, or the lower envelope.  It is unclear which, if any, smoothing procedure would give a meaningful light curve.  (3) Active state data typically include many flaring events, which would produce spurious features in any fitting or smoothing algorithm.  For example, in Figure 2c of LHO98, there is a ~.2 magnitude flare apparent near phase 0.5, with smaller flares apparent near phases 0.2 and 0.3; the active light curves of CMBO08 contain many such flares.  Though large flares could be identified and removed, smaller flares are so numerous that they could not be identified individually and would have a major impact on the light curve shape.

Small flares are present even in passive data.  I8 shows a clear flare: Though its residual in Figure 2c is almost perfectly flat, there are three consecutive points near phase 0.25 which peak 0.04 mag above the fit.  Curve I4 similarly has a flat residual with one point 0.05 mag above all the rest.  These features are so isolated and peaked that it is easy to identify them as minor flares, separate from the smooth variation due to the changing viewing angle of the disk.  Such small flares could not be identified in the SMARTS data, which have larger photometric errors and just one exposure per night.  It is therefore likely that many such flares exist within the SMARTS passive data.  We believe this is the origin of the uneven residuals given by fits to the SMARTS curves V6 I6 and H6 in Figure 2: averaging over many minor flares would lead to spurious features, which would appear as uneven residuals to the fit.  If the flares are more prominent at some phases than others, they would naturally produce the phase-modulated residuals we see.  This scenario seems likely, given that the only curves to show systematic residuals are those which average over a period of years. 

Our results put A0620 in good agreement with the results of Bailyn et al.\ (1998), who find that the masses of black holes in transient low-mass X-ray binaries are strongly clustered near 7M$_\odot$.  In that paper, the mass of A0620's black hole was only constrained to lie between 4 and 25M$_\odot$.  It is therefore remarkable that our determination of $i$, along with refined determinations of the mass function and mass ratio, put A0620 in such good agreement with the masses of the other objects discussed in Bailyn et al.\ (1998).  As discussed there, the clustering of masses at such a large value is unexpected: simple models of stellar evolution would predict that lower-mass objects would be more common and certainly do not predict clustering at any particular mass.  Our result thus strengthens the conclusion of Bailyn et al.\ (1998) and deepens the question of why black holes should preferentially form at a specific mass, especially one much larger than the 3M$_\odot$ lower limit for the formation of stellar-mass black holes.

\section{Conclusions}
1. Previous disagreements on the inclination of A0620 have likely resulted from three effects: (1) state changes were not recognized before CBMO08, and non-passive data were often included in curves to be fit for the determination of $i$, (2) the light curve may change shape from night-to night, and combining curves of different shapes can give misleading results, and (3) the disk contribution has not been adequately constrained in previous work.

2. After correcting for the three effects above, we find that the data consistently point to a single value of $i$.  Based on fits to many light curves which had previously given conflicting results, we find $i=50\fdg98\pm0.87$, implying $M=6.61\pm0.25M_\odot$ and $M_2=0.40\pm0.045M_\odot$.

3. Based on our firm dynamical model and zero-disk magnitudes of A0620, we estimate the distance to A0620 to be $d=1.06\pm0.12$kpc.

4. We find that all non-ellipsoidal variations are dominated by disk light: the disk dominates secular changes in brightness, the flickering present in non-passive states, and periodic modulations in the shape of the light curve.   The disk fraction is greater than 10\% in all curves we study, including passive state, infrared light curves.

5. The lessons learned from A0620 may help make robust determinations of $i$ in other objects.  Specifically, we have shown the importance of states, disk light, and night-to-night changes in curve shape.  These effects are probably not unique to A0620 and our techniques for handling them may lead to more robust determinations of $i$ in other objects.

\acknowledgements 
This work was supported by NSF Graduate Research Fellowship DGE-0202738 to AGC and NSF/AST grants 0407063 and 0707627 to CDB.

\end{document}